\documentclass[aps,prb,twocolumn,letterpaper,superscriptaddress,showpacs]{revtex4-1}
\usepackage{graphicx}
\usepackage{CJK}
\usepackage{amsmath}
\usepackage{color}
\usepackage{dcolumn}% Align table columns on decimal point
\usepackage{bm}% bold math
\usepackage{hyperref}% add hypertext capabilities
\usepackage{txfonts}

\newcommand{\lfco}{LiFeCr$_{4}$O$_{8}$}
\newcommand{\lgco}{LiGaCr$_{4}$O$_{8}$}
\newcommand{\lico}{LiInCr$_{4}$O$_{8}$}

\begin{document}
\begin{CJK*}{UTF8}{bsmi}
\title{
M\"ossbauer spectroscopy study of the magnetostructural and spin-state transitions \\ in the breathing pyrochlore \lfco\
}
\author{Bo Zhang}
\author{Wei Ren}
\author{Shengyu Yang}
\affiliation{Key Lab for Magnetism and Magnetic Materials of the Ministry of Education, School of Physical Science and Technology, Lanzhou University, Lanzhou 730000, China}
\author{Qifeng Kuang}
\author{Da Li}
\affiliation{Shenyang National Lab for Materials Science, Institute of Metal Research, Chinese Academy of Sciences, and School of Materials Science and Engineering, University of Science and Technology of China, 72 Wenhua Road, Shenyang, 110016, China}
\author{Xin Liu}
\affiliation{School of Resources and Materials, Northeastern University at Qinhuangdao, Qinhuangdao 066004, PR China}
\affiliation{School of Materials Science and Engineering, Northeastern University, Shenyang 110819, PR China}
%\author{Tinghai Zhang}
%\author{Yiwen Dong}
\author{Anmin Zhang}
\author{Hua Pang}
\author{Liyun Tang}
\author{Liang Qiao}
\author{Fashen Li}
\author{Zhiwei Li}\email{zweili@lzu.edu.cn}
\affiliation{Key Lab for Magnetism and Magnetic Materials of the Ministry of Education, School of Physical Science and Technology, Lanzhou University, Lanzhou 730000, China}
% \affiliation{School of Physical Science and Technology, Lanzhou University, Lanzhou 730000, China}

\date{\today}

\begin{abstract}
We report on investigations of the complex magnetostructural and spin-state transitions in the breathing pyrochlore \lfco\ by means of magnetization, M\"ossbauer spectroscopy, and density functional theory (DFT) calculations.
Three transitions corresponding to the ferrimagnetic transition at $T_N\sim94$\,K, the spin-gap transition at $T_{SG}\sim50$\,K, and the magnetostructural transition at $T_{MS}\sim19$\,K were observed from the $\chi$(T) curve, whereas only $T_N$ and $T_{MS}$ were evidenced for the Fe site from our M\"ossbauer measurements, suggesting that the spin-gap transition is absent at the Fe site.
This indicates that the spin-gap transition is an effect of the breathing Cr$_4$ lattice, in agreement with our DFT calculations from which we see nearly decoupled electronic states for the FeO$_4$ and CrO$_6$ units.
From the temperature dependence of the hyperfine magnetic field we also observed a spin-state transition for the Fe spins at $T_{MS}$ consistent with earlier neutron diffraction measurements.
These local characteristics are believed to be important for a complete understanding of the complex magnetostructural coupling effects observed in similar systems.
\end{abstract}


\maketitle
\end{CJK*}

\section{Introduction}
\begin{figure*}[th]
\includegraphics[width=1.8\columnwidth,clip=true]{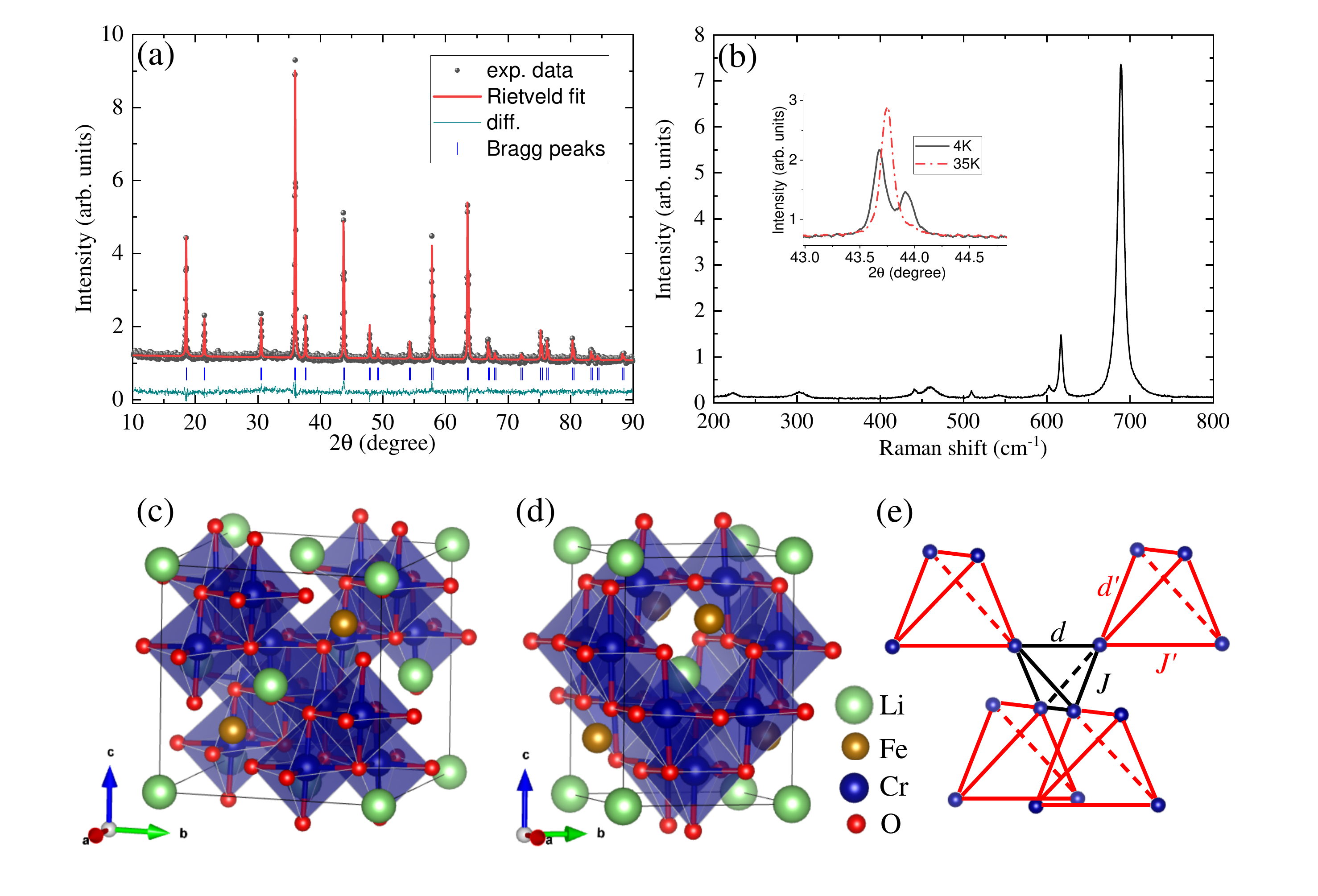}
\caption{\label{fig:XRD}
(color online)
(a) Rietveld refinement of the room temperature X-ray powder diffraction data of \lfco. (b) Raman spectrum of \lfco\ at room temperature. The inset shows the splitting of the (400) XPRD peak at 35\,K into the (220) and (004) peaks at 4\,K well below the tetragonal transition. (c) Cubic ($F\bar{4}3m$) and (d) tetragonal ($I\bar{4}m2$) crystal structures. (e) Schematic representation of the breathing pyrochlore lattice of \lfco\ where the distortion ($d'/d$) on the Cr$_4$ network has been exaggerated for better view. The crystal structures were drawn by using the software VESTA \cite{vesta}.
}\end{figure*}

Geometrically frustrated magnetic systems have been an interesting playground for condensed matter physicists over the last 30 years \cite{rmp.82.53}.
The chromium spinels with a commen formula of $A$Cr$_2$$X_4$, where $A$ is usually nonmagnetic atoms and $X$ stands for O, S, and Se atoms, are a rich family of such compounds that exhibit various interesting phenomena, such as the zero-energy excitation mode, heavy fermionic behavior, spin lattice coupling, and field induced transitions \cite{jpsj.79.011004}.
More interestingly, when two different types of elements are put at the $A$ site, it leads to the formation of the breathing lattice with alternating large and small Cr$_4$ tetrahedrons due to the ordering of the two $A$-site ions \cite{prl.110.097203, prb.106.024407} (see Fig.\ref{fig:XRD} (e) for an illustration). This type of ordering can minimize the electrostatic energy arising from the large difference in the valence states between the two $A$-site ions (e.g., Li$^+$ vs Ga$^{3+}$/In$^{3+}$ \cite{prl.110.097203}).
It was found theoretically that the breathing lattice may host the hedgehog spin textures (magnetic monopoles) when the third nearest-neighbor (NN) exchange interaction $J_3$ is large enough and if the alternating NN exchange interactions are different, $J_1\neq J_1'$ \cite{prb.103.014406, prb.106.064412}.
This emphasizes the importance of the magnetic interactions between the Cr spins.

The magnetic interactions are dominated by anti-ferromagnetic (AFM) correlations between Cr ($S=3/2$) spins in Li(Ga,In)Cr$_4$O$_8$ with considerably reduced magnetic moments and AFM transition temperatures, suggesting that frustration also plays an important role \cite{prl.110.097203,prl.127.147205,prl.127.147205,jpsj.84.043707}. The magnetic properties also depend significantly on the so-called breathing factor $B_f = J'/J$ since the interaction between nearest-neighbor Cr atoms is distance sensitive \cite{prb.77.115106,prl.125.167201,npj.4.63}. For example, the Ga-based sample exhibits AFM short-range order below $\sim$45\,K like conventional Cr spinel oxides, while the In-based compound shows spin-gap behavior below about 65\,K \cite{prl.110.097203}. The structural and magnetic properties were further investigated by substitution of the Cr atoms by other elements on the breathing lattice \cite{pssb.257.1900685,rp.35.105379,cpl.33.127501} or by application of external magnetic field \cite{prb.101.054434,jpsj.91.023710}.

Moreover, it is also very interesting to replace one of the nonmagnetic $A$-site ions by a magnetic one, which can introduce further magnetic interactions between $A$ and Cr spins. For example, R. Saha et. al. have studied \lfco\ and found interesting magnetoelectric effects with multi-magnetic phase transitions, namely, a ferrimagnetic transition at $T_N\sim94$\,K, a spin-gap transition at $T_{SG}\sim60$\,K, and a magnetostructural transition at $T_{MS}\sim23$\,K where the high temperature collinear magnetic structure changes to a low temperature conical magnetic structure \cite{prb.96.214439}. Recently, large magnetic-field-induced strain at $T_{MS}$ was reported by Y. Okamoto et. al. \cite{jpsj.91.023710} indicating strong spin-lattice coupling in this compound. Surprisingly, however, the Fe$_4$ tetrahedrons exhibit larger volumes for the low temperature conical magnetic phase than the high temperature ferrimagnetic phase \cite{prb.96.214439} even though a large volume contraction was observed when lowering the temperature through the magnetostructural transition \cite{jpsj.91.023710}.

To better understand the physics behind these interesting phenomena, a local probe study at the local Fe site becomes important.
Therefore, we investigated the title compound \lfco\ by using M\"ossbauer spectroscopy which is only sensitive to the $^{57}$Fe ions at the $A$-site.
From the temperature dependence of the hyperfine parameters, we confirmed the spin-state transition of the Fe spins and we also provide evidence that the spin-gap transition, which is observed from our static and dynamic magnetic measurements, is absent at the Fe site suggesting that the spin-gap transition is only an effect of the breathing Cr$_4$ lattice. These results can be understood with our density functional theory (DFT) calculations where we see nearly decoupled electronic states for the FeO$_4$ and CrO$_6$ units.
These local properties provide a better understanding of the complex magnetostructural coupling effects observed in this system.

\section{Experiments}

%\textit{Experimental detail:}
A polycrystalline powder sample of \lfco\ was synthesized by using the conventional solid state reaction technique \cite{prb.96.214439}. Stoichiometric amounts of Li$_2$CO$_3$, Fe$_2$O$_3$ and Cr$_2$O$_3$ (all from Alfa Aesar, 99.99\%) were thoroughly mixed, pelletized and heated in air at 1050\,$^o$C for 15\,h with a cooling rate of 50\,$^o$C/h to room temperature. The homogenization and heating procedure were repeated several times to improve the sample quality.
Phase purity was checked by X-ray powder diffraction (XRPD) with Cu K$_{\alpha}$ radiation using a X'Pert Pro X-ray diffractometer (Philips, Netherlands) and the data refinement was done by using the FullProf suite \cite{fullprof}. The Raman measurement was performed in a confocal back-scattering geometry using a Jobin Yvon LabRAM HR Evolution spectrometer equipped with a 1800 lines/mm grating, a liquid-nitrogen-cooled back-illuminated charge-coupled device detector and a 532\,nm laser.
Static magnetic measurements were carried out with a dc superconducting quantum interference device (SQUID) magnetometer (Quantum Design) in the temperature range of 2$\sim$300\,K. The dynamic magnetic properties, the real and imaginary parts of the complex magnetic susceptibility, were measured by means of a precision LCR Meter (HP4284A) with a cryostat in the temperature range of 4$\sim$300\,K at a frequency of 138\,kHz.
M\"ossbauer measurements were performed in transmission geometry with a conventional spectrometer working in constant acceleration mode. A 50\,mCi $\gamma$-ray source of $^{57}$Co embedded in Rh matrix and vibrating at room temperature was used. The drive velocity was calibrated by using an $\alpha$-Fe foil. The isomer shift quoted in this work are relative to that of the $\alpha$-Fe at room temperature.

The computational work was carried out by using the \textsc{ELK} code \cite{elk}, which is based on the full potential linearized augmented plane waves (FP-LAPW) method. The Perdew-Wang/Ceperley-Alder local spin density approximation (LSDA) exchange-correlation functional \cite{prb.45.13244} was used. LSDA+U calculation was done in the fully localized limit (FLL) and by means of the Yukawa potential method \cite{prb.52.R5467} with a screening length of $\lambda=2.0$ (other $\lambda$ values give similar results, and thus were not discussed) for both Fe and Cr $d$ electrons. Slater integrals are calculated according to $\lambda$ and the resulting Coulomb interaction parameters are $U=5.35$\,eV and $J=1.13$\,eV for Fe and $U=4.00$\,eV and $J=0.94$\,eV for Cr, respectively. The muffin-tin radii $R_{MT}$ were set automatically by \textsc{ELK} to 1.80\,a.u., 2.0313\,a.u., 2.2248\,a.u. and 1.4122\,a.u. for Li, Fe, Cr, and O atoms, respectively.
For the nonmagnetic, ferromagnetic, and ferrimagnetic states calculations, the plane-wave cutoff was set to $R_{MT}\times |\textbf{G}+k|_{max} = 7.0$ and the maximum \textbf{G}-vector for the potential and density was set to $|\textbf{G}|_{max}=12.0$. A $\textbf{k}$-point mesh of $8\times8\times8$ was used and the spin orbital coupling (SOC) was not considered.
For the conical magnetic state calculations, the plane-wave cutoff was increased to $R_{MT}\times |\textbf{G}+k|_{max} = 8.5$ and the maximum \textbf{G}-vector was set to $|\textbf{G}|_{max}=14.0$. A reduced $\textbf{k}$-point mesh of $4\times4\times2$ (total of 32 k-points) was used to speed up our calculations since SOC was included in these calculations. Experimental lattice parameters of \lfco\ at 298\,K (space group: $F\bar{4}3m$, $a=2.2779$\,{\AA}) and 3.5\,K (space group: $I\bar{4}m2$, $a = 5.85755$\,{\AA} and $c = 8.24301$\,{\AA}) were taken from Ref.\cite{prb.96.214439} for our collinear and conical magnetic structure calculations, respectively, without further optimization.

\section{Results and discussion}
To check the crystal structure and sample quality, we performed room temperature XRPD measurements as shown in Fig. \ref{fig:XRD} (a) together with Rietveld refinement. The Rietveld analysis confirms the noncentrosymmetric $F\bar{4}3m$ space group of the \lfco\ compound and the determined lattice parameter $a=8.2764 (3)\,{\AA}$ as being consistent with previously reported values of $8.27779(1)\,{\AA}$ \cite{prb.96.214439} and $8.2753(3)\,{\AA}$ \cite{jpsj.91.023710}. In this structure, Li$^+$ and Fe$^{3+}$ ions present in $4a$ and $4d$ Wyckoff sites, respectively, and the ordering between these two ions results in a different amount of chemical pressure on the pyrochlore network of Cr$_4$ leading to the so called breathing pyrochlore lattice \cite{prb.96.214439, prl.110.097203}. The sample quality was also evidenced by the sharp Raman peaks shown in Fig. \ref{fig:XRD} (b) which exhibits the same pattern as an earlier report \cite{prb.96.214439}. The low temperature tetragonal transition \cite{prb.96.214439} was shown by the splitting of the (400) XPRD peak measured at 35\,K into the (220) and (004) peaks at 4\,K as shown in the inset of Fig. \ref{fig:XRD} (b). The schematic representation of the crystal structure and the Cr$_4$ breathing pyrochlore lattice are shown in Fig. \ref{fig:XRD} (c) - (e), where the magnitude of the breathing has been exaggerated for better visual effect.

\begin{figure}[th]
\includegraphics[width=1.0\columnwidth,clip=true]{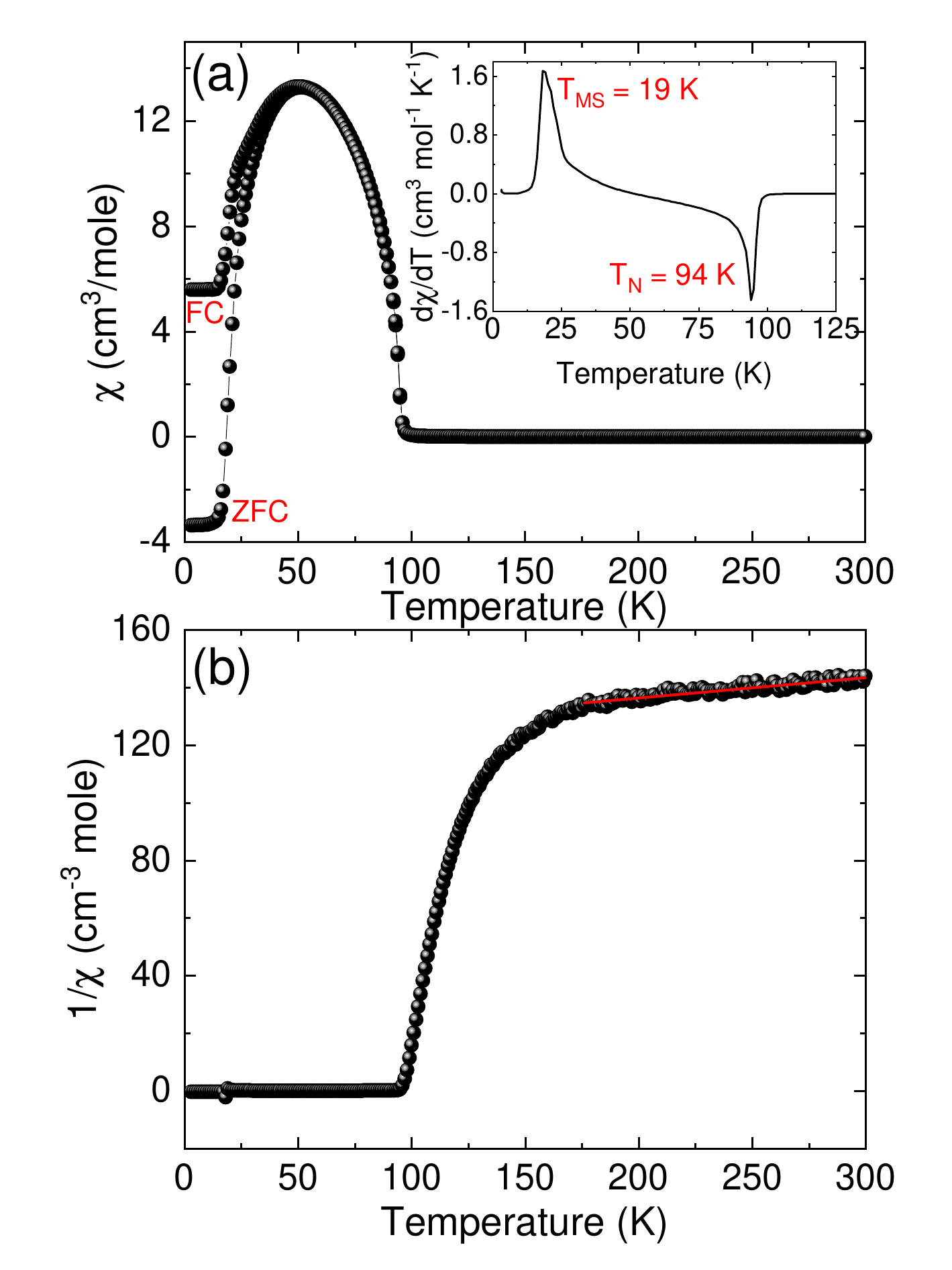}
\caption{\label{fig:MT}
(color online)
(a) Temperature dependence of the magnetic susceptibility, $\chi$(T), of \lfco\ measured in ZFC-FC mode with $H=100\,Oe$. Inset: enlargement of the first-order derivative of the susceptibility with respect to measured temperature plotted against temperature to show the ferrimagnetic, $T_N=94\,K$, and low temperature magnetostructural, $T_{MS}=19\,K$, transitions. (b) Temperature dependence of inverse susceptibility of \lfco. The red solid line between 175 and 300\,K is the Curie-Weiss fit to the high temperature experimental data.
}\end{figure}

In Fig. \ref{fig:MT} (a), we present the magnetic susceptibility, $\chi$(T), as a function of temperature measured with an applied magnetic field of $H=100\,Oe$ in both zero-field-cooled (ZFC) and field-cooled (FC) modes. The overall behavior of the $\chi$(T) data is similar to earlier reports \cite{prb.96.214439,jecs.27.903} only with a negative initial value at low temperatures in the ZFC measurement, which is due to the measurement history effect where the initial net magnetization aligns antiparallel with the applied magnetic field at the starting point.
This is often observed in systems with two or more magnetic sublattices showing an antiparallel ordering (ferrimagnetism in \lfco) with different temperature dependencies of their magnetization below the ordering temperature \cite{prep.556.1}.
The first-order derivative of the susceptibility with respect to measured temperature, as shown in the inset of Fig. \ref{fig:MT} (a), was used to extract the ferrimagnetic, $T_N=94\,K$, and low temperature magnetostructural, $T_{MS}=19\,K$, transitions as were reported also by earlier works \cite{prb.96.214439, jpsj.91.023710}.
The initial decrease of $\chi(T)$ with decreasing temperature at around $T_{SG}\sim50\,K$ is related to the spin-gap transition arising from the breathing distortion in similar compounds \cite{prb.96.214439,prl.110.097203,prb.94.064420,prb.91.174435}.
The high temperature data were analyzed by fitting the Curie-Weiss law, $\chi = C/(T - \theta)$, to the linear region of the inverse susceptibility data as shown in Fig. \ref{fig:MT} (b). The Curie constant $C$ of a system of $N$ spins S can be expressed as
\begin{equation}
\label{eqCW}
C = \frac{Ng^2S(S+1)\mu_B^2}{3k_B},
\end{equation}
where $\mu_B$ is the Bohr magnetron and $k_B$ is the Boltzmann's constant. For a system containing different spin values, we can replace $S(S+1)$ in equation (\ref{eqCW}) by its average value $\langle S(S+1)\rangle$ in the mean field approximation. Then, for the title compound \lfco, if we assume $S=5/2$ and $S=3/2$ for Fe$^{3+}$ and Cr$^{3+}$ spins, respectively, one obtains the theoretical value of $9.75\,\mu_B/f.u.$. However, from the fitted value of $C$, we obtained an effective magnetic moment of $\mu_{eff} = 10.5(2)\,\mu_B$, slightly larger than the theoretical value, but close to the earlier reported value of $\mu_{eff} = 10.69\,\mu_B$ \cite{prb.96.214439}. Strong geometrical frustration is indicated by the large frustration index ($f=|\theta|/T_N=18$) obtained from the paramagnetic intercept. Similarly, large values of $f=12$ for \lfco\ \cite{prb.96.214439}, $f=21$ for \lico\ and $f=47$ for \lgco\ have been reported \cite{prl.110.097203}.

\begin{figure}[th]
\includegraphics[width=1.0\columnwidth,clip=true]{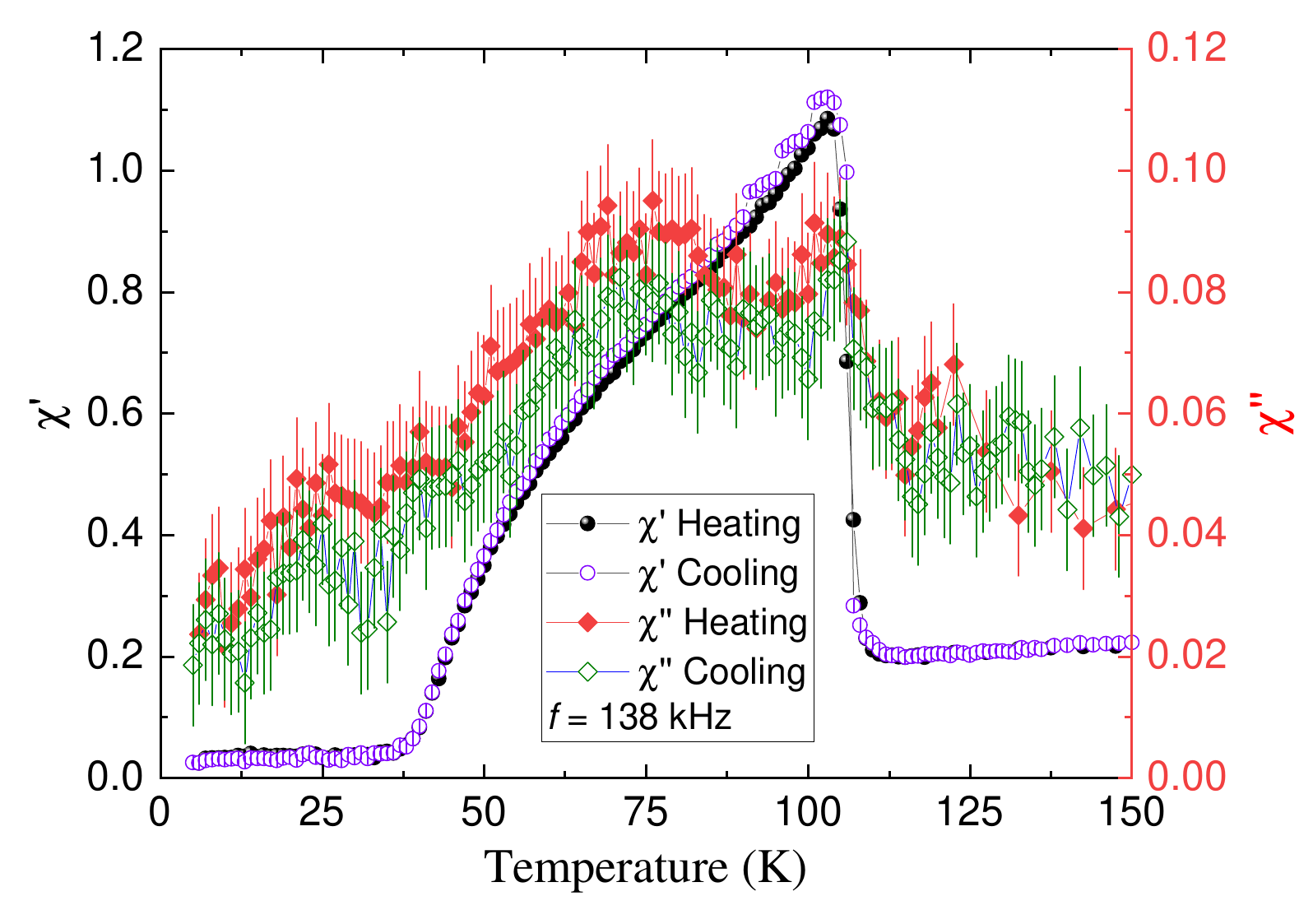}
\caption{\label{fig:muT}
(color online)
Temperature dependence of the complex magnetic susceptibility recorded at 138\,kHz with both cooling and heating procedures.
}\end{figure}

\begin{figure}[th]
\includegraphics[width=0.95\columnwidth,clip=true]{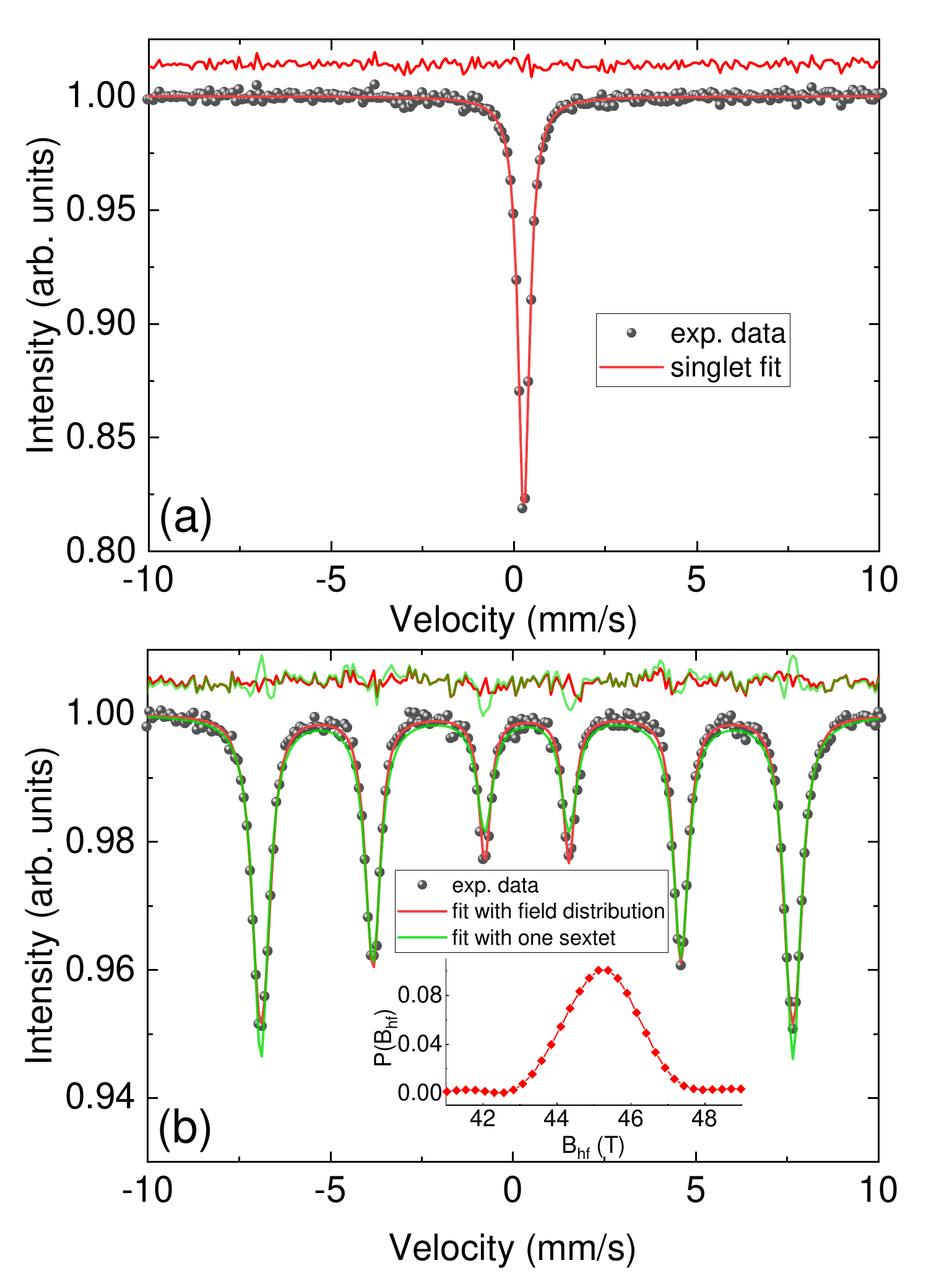}
\caption{\label{fig:moss}
(color online)
(a) Room temperature M\"ossbauer spectrum of \lfco\ (dots) with singlet fit (red solid line). (b) Low temperature M\"ossbauer spectrum of \lfco\ taken at 4.2\,K (dots) together with two different fitting procedures shown as red (field distribution) and green (single sextet) solid lines, as discussed in the text. Inset shows the field distribution corresponding to the red line fit in the main figure. The difference between the experimental data and the calculated data is also shown above the spectra.
}\end{figure}

To clarify the elusive spin-gap transition at $T_{SG}\sim50\,K$ shown in the $\chi(T)$ curve, we made temperature dependence measurements of the complex magnetic susceptibility at 138\,kHz as shown in Fig. \ref{fig:muT}. Two transitions corresponding to $T_N \sim 105$\,K and $T_{MS} \sim 35$\,K can be seen from the real part, $\chi'(T)$, of the complex magnetic susceptibility. Interestingly, the spin-gap transition $T_{SG} \sim 75$\,K can be clearly seen as a broad peak from the imaginary part, $\chi''(T)$, of the complex susceptibility. However, we would like to note that these transitions extracted from the dynamic magnetic susceptibility are higher than that determined from the above static measurements. The reason for this might be twofold: I) the transition temperature might be frequency dependent as observed in other systems \cite{prb.58.14937, jmmm.483.178} and II) short range magnetic correlations might be already exist well above the static magnetic transitions which was captured by our dynamic measurements. The latter is consistent with the slow volume change with decreasing temperature well above $T_N$ due to the strong magnetostructural coupling effect \cite{prb.96.214439,jpsj.91.023710}.

\begin{figure*}[th]
\includegraphics[width=2.0\columnwidth,clip=true]{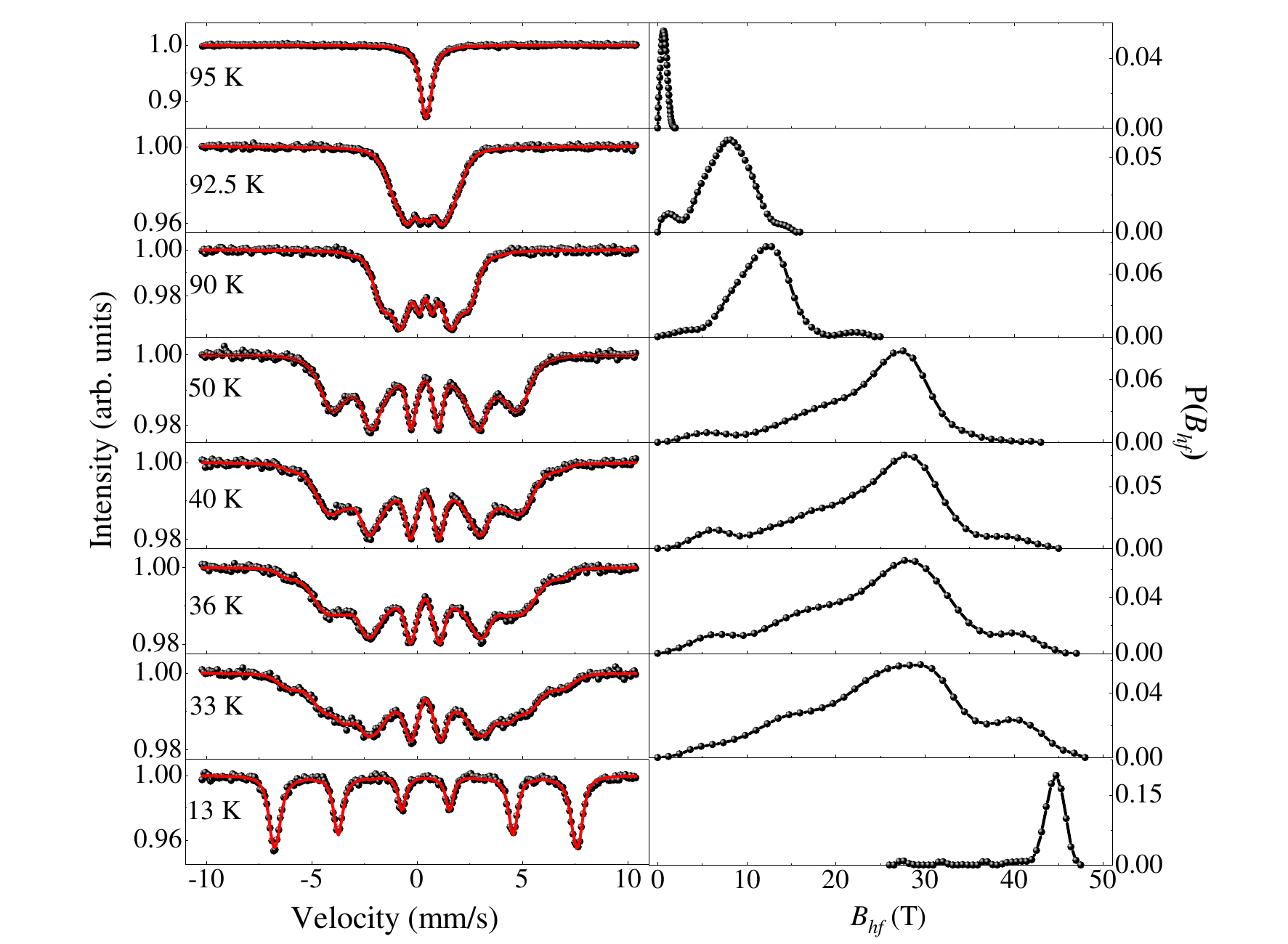}
\caption{\label{fig:mossdist}
(color online)
 (Left panel) M\"ossbauer spectra of \lfco\ (dots) taken in the temperature range of 13\,K$\sim$95\,K together with theoretical fits (red lines) using the hyperfine magnetic field distribution model. (Right panel) The corresponding field distribution profiles of the theoretical fits shown in the left panel. The measurement temperatures of these spectra are also indicated.
}\end{figure*}

To investigate the local properties at the Fe site, we made $^{57}$Fe M\"ossbauer spectroscopy measurements.
Fig. \ref{fig:moss} (a) presents the room temperature M\"ossbauer spectrum of \lfco\ with a singlet fit to the experimental data. The obtained isomer shift is $\delta(RT) = 0.267(1)$\,mm/s. We also tried to model the data with a doublet, however, it results in an effectively zero quadruple splitting, indicating the absence of the electric field gradient at the iron site. These results agrees well with the picture of the Fe$^{3+}$ ion sitting at the tetrahedron site. The fitted spectral line width has a relatively small value of $\sim0.363(3)$\,mm/s, only slightly larger than the value of the standard sample $\sim0.326(4)$\,mm/s, suggesting a unique local environment of the Fe$^{3+}$ ions which is consistent with the crystal structure of \lfco\ with the $A$-site ordering of Li$^{+}$ and Fe$^{3+}$ ions.

Fig. \ref{fig:moss} (b) shows the M\"ossbauer spectrum taken at 4.2\,K. We tried to fit the spectrum with one sextet (solid green line) as shown in the figure.
Larger $\chi^2=3.04$ was obtained from this fit, suggesting a bad agreement between the fit and the experimental data, which can also be seen from the green difference curve shown above the spectrum. Considering the low temperature conical magnetic structure \cite{prb.96.214439}, a small hyperfine magnetic field distribution might be expected if an anisotropic hyperfine coupling tensor (\textbf{A}) were assumed ($B_{hf}\propto \textbf{A}\cdot \textbf{S}$)  \cite{prb.97.104415}. Magnetic field distribution may also be caused by nanosized conical magnetic domains as indicated by the broadening of the magnetic reflections from neutron diffraction, $\sim 62$\,{\AA} \cite{prb.96.214439}. Therefore, we modeled the spectrum at 4.2\,K with a magnetic field distribution (red solid line) as shown in the figure. Considerable improvement of the fit has been obtained by the field distribution model, $\chi^2=1.25$ (also see the red line difference curve).
However, one should note that we can not exclude other possible reasons for the observed magnetic field distribution.
The determined isomer shift is $\delta(4.2\,K) = 0.379(1)$\,mm/s and the average magnetic field amounts to $\langle B_{hf}\rangle(4.2\,K) = 45.3$\,T. The fitted quadruple splitting is almost zero, $\sim0.002(3)$\,mm/s, indicating that the local symmetry of the FeO$_4$ tetrahedron has not been affected much by the magnetostructural distortion at $T_{MS} = 19$\,K. This agrees well with the Rietveld refined crystallographic data, where two equal sets of Fe-O bond length were obtained for the FeO$_4$ tetrahedron \cite{prb.96.214439} for the low temperature tetragonal phase.

Fig. \ref{fig:mossdist} shows the M\"ossbauer spectra (dots) taken in the temperature range of 13\,K$\sim$95\,K together with theoretical fits (red lines) using the hyperfine magnetic field distribution model and the corresponding field distribution profiles are shown in the right panel. Clearly, the spectra taken in the temperature range between the ferrimagnetic transition ($T_N=94$\,K) and magnetostructural transition ($T_{MS}=19$\,K) exhibit much broader spectral line width, suggesting a much wider distribution of the magnetic field than that for the 4.2\,K spectrum as shown in Fig. \ref{fig:moss} (b). This wide distribution effect can be attributed to the mixing of the low temperature tetragonal phase with conical magnetic structure and the intermediate temperature cubic phase with collinear magnetic structure, which was further corroborated with the multi-component nature of the field distribution profiles shown on the right panel of Fig.\ref{fig:mossdist}. This two phase mixing phenomenon due to the first order magnetostructural transition has been reported earlier by neutron diffraction \cite{prb.96.214439} and magnetic field induced strain measurements \cite{jpsj.91.023710}. This is also a very common phenomena for other similar breathing pyrochlore chromate spinels \cite{prl.110.097203, prb.91.174435, prl.127.147205}.

\begin{figure}[th]
\includegraphics[width=1\columnwidth,clip=true]{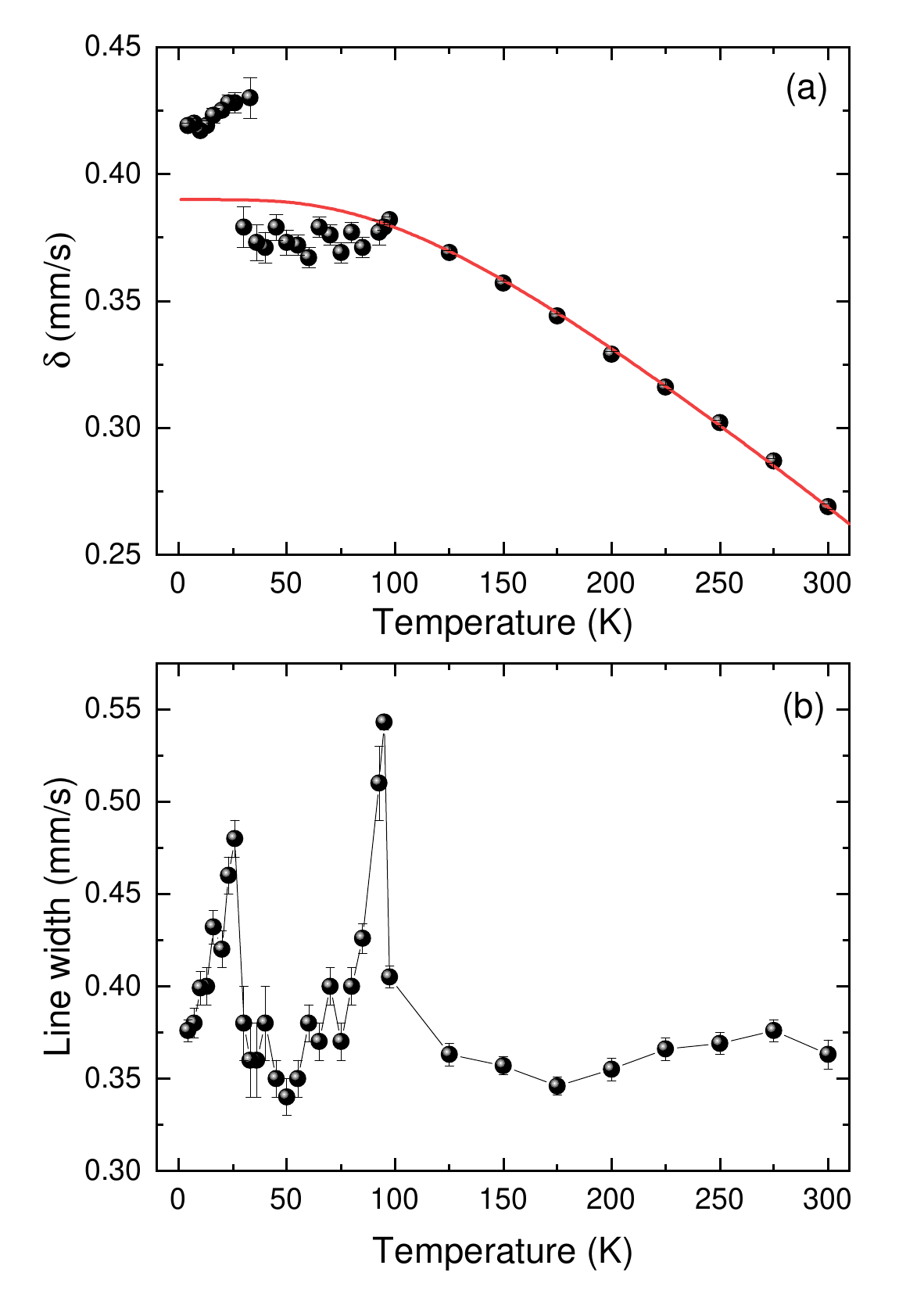}
\caption{\label{fig:IS}
(color online)
Temperature dependence of the fitted M\"ossbauer hyperfine parameters (a) isomer shift $\delta(T)$ and (b) spectral line width. Solid line in (a) is theoretical fit to the data using the Debye model and solid line in (b) is a guide to the eye.
}\end{figure}

The isomer shift $\delta(T)$ and spectral line width, determined from the fits shown in Fig.\ref{fig:mossdist}, are shown as a function of temperature in Fig.\ref{fig:IS} (a) and (b), respectively. Two anomalies that correspond to the ferrimagnetic and magnetostructural transitions can be seen at $T_N = 94$\,K and $T_{MS} \sim 27$\,K. The magnetostructural transition temperature determined from our M\"ossbauer measurements is a little higher than that of our susceptibility measurement but agrees well with that reported in previous studies $\sim23$\,K \cite{prb.96.214439} and $\sim30$\,K \cite{jpsj.91.023710}. The small difference can be caused by different measurement techniques or the detailed method used in determining the transition temperature and small difference in the stoichiometry of the different samples used in different works.
The red solid line shown in Fig.\ref{fig:IS} (a) is a theoretical fit to the experimental data in the high temperature range by using the Debye model.
In the Debye model, the temperature dependence of $\delta(T)$ is expressed by the following equation \cite{PGmbook2011}
\begin{equation}
\begin{split}
\delta(T) = \delta(0) - \frac{9}{2}\frac{k_BT}{Mc}(\frac{T}{\Theta_D})^3\int_0^{\Theta_D/T}\frac{x^3dx}{e^x-1}
\end{split}
 \label{eqIS}
\end{equation}
where $\delta(0)$ is the temperature independent chemical shift, and the second part is the temperature dependent second-order Doppler shift. Here, $k_B$ is the Boltzmann constant, $M$ is the mass of the M\"ossbauer nucleus, $c$ is the speed of light and $\Theta_D$ is the Debye temperature.
The determined Debye temperature is $\Theta_D=443(8)$\,K and the temperature independent chemical shift is $\delta(0)=0.380(5)$\,mm/s.

\begin{figure}[th]
\includegraphics[width=1\columnwidth,clip=true]{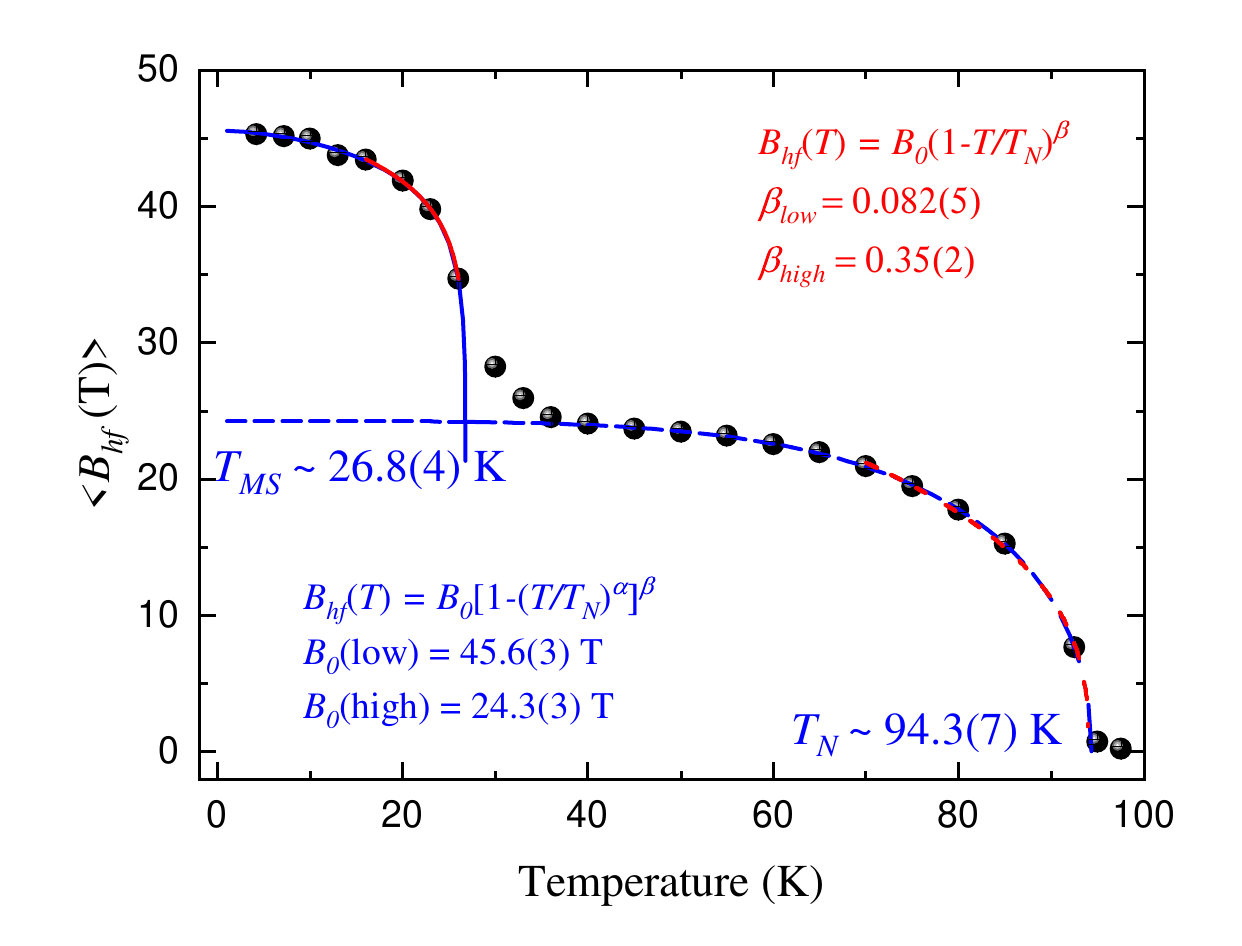}
\caption{\label{fig:Bhf}
(color online)
Temperature dependence of the average hyperfine magnetic field $\langle B_{hf}(T)\rangle$ for \lfco. Power-law, $B_{hf}(T) = B_{0}(1 - T/T_N)^{\beta}$, fits to the experimental data close to the transition region were used to estimate the critical exponent $\beta$ (see the red lines). To determine the saturation value of the hyperfine magnetic field and the transition temperature more accurately, a different kind of power law $B_{hf}(T) = B_{0}[1 - (T/T_N)^{\alpha}]^{\beta}$ was used to fit the experimental data in a wider temperature range (see the blue lines). Solid lines and dashed lines shown in the figure correspond to the low and high temperature fits, respectively.
}\end{figure}

The temperature dependence of the average hyperfine magnetic field $\langle B_{hf}(T)\rangle$ as a function of temperature is shown in Fig.\ref{fig:Bhf}. Like other hyperfine parameters, two transitions at $T_N=94.3(7)$\,K and $T_{MS}=26.8(4)$\,K can be seen. Power law $B_{hf}(T) = B_{0}[1 - (T/T_N)^{\alpha}]^{\beta}$ was used to fit the experimental data in a wider temperature range to, more accurately, determine these transition temperatures and the saturation hyperfine magnetic field, $B_{0}$, for the high temperature ferrimagnetic state $B_0(high)=24.3(3)$\,T and the low temperature conical magnetic state $B_0(low)=45.6(3)$\,T.
However, to study the critical behavior of \lfco, the usual power law function $B_{hf}(T) = B_{0}(1 - T/T_N)^{\beta}$ was used to fit the experimental data close to the transition temperature range. The determined critical exponents are $\beta_{high} = 0.35(2)$ and $\beta_{low} = 0.082(5)$ for the ferrimagnetic state and conical magnetic state, respectively. The value of $\beta_{high} = 0.35(2)$ can be identified with the three dimensional Heisenberg critical exponent in view of the cubic symmetry of \lfco\ ($\beta \sim 0.36$ \cite{prl.39.95,prb.21.3976}). On the other hand, the low temperature critical exponent has a value of $\beta_{low} = 0.082(5)$ which is close to the theoretical value ($\beta = 1/12$) of the ($q=4$)-state Potts model which is the limit of a sequence of models ($q>4$) with a discontinuous, first-order transition \cite{jpa.15.3329, jpcm.20.275233}.
This is consistent with the observed first-order magnetostructural transition from the high temperature collinear ferrimagnetic phase to the low temperature conical magnetic phase \cite{prb.96.214439}.
However, since only a few temperature points were available in our fitting, the observed crossover of the critical behavior needs further investigation with more measurements in the temperature range near the transition point.

\begin{figure*}[th]
\includegraphics[width=2.0\columnwidth,clip=true]{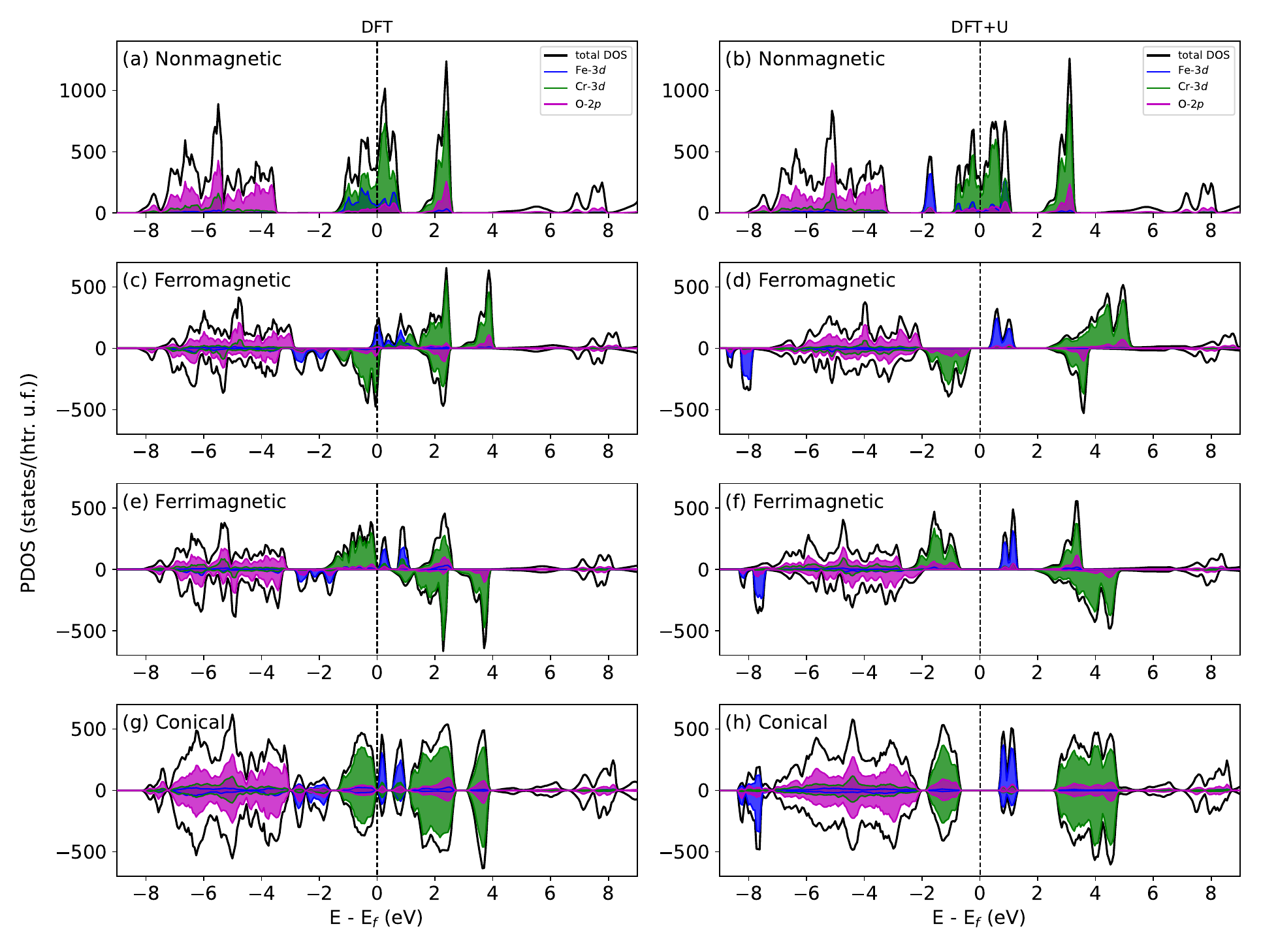}
\caption{\label{fig:dos}
(color online)
 Total and partial densities of states of \lfco\ for nonmagnetic, ferromagnetic, ferrimagnetic and the conical magnetic states calculated with DFT (left panel) and DFT+U methods (right pannel).
 Spin orbital coupling was only included for the conical magnetic state calculations. Please see the text for details.
}\end{figure*}

To understand the microscopic origin of the observed experimental results, we made DFT calculations using the Elk code.
The total and partial densities of states of \lfco\ that correspond to nonmagnetic, ferromagnetic, ferrimagnetic and the conical magnetic states are shown in Fig.\ref{fig:dos} (a)-(h).
For the ferro-/ferri- magnetic structures, the spins are all along the crystal $c$-axis and the Fe spins are aligned antiparallel with the Cr spins in the ferrimagnetic structure. For the non-collinear conical magnetic structure, the spin directions are fixed to the values taken from Ref\cite{prb.96.214439} with an incommensurate propagation vector $\textbf{k} = (1/2, \delta = 0.4383, 1/2)$, and the magnitude of the spins for both Fe and Cr atoms were allowed to change.
It is clear that our DFT solutions result in metallic ground states for the nonmagnetic and ferromagnetic calculations and only soft gaps with minimum densities of states close to zero at the Fermi level were opened for the ferrimagnetic and conical magnetic calculations.
These results were opposed to the experimentally observed insulating behavior since DFT calculations usually underestimate the Coulomb correlation effects among the 3$d$ electrons, which is often found to be responsible for the insulating behavior of transition metal oxides \cite{jpcm.9.767,prb.78.045120}.
The nonmagnetic calculation with the DFT+U method also gives a metallic state as seen from Fig.\ref{fig:dos} (b), whereas the magnetic states provide the insulating ground state with a hard gap of $\sim$0.65\,eV for the ferromagnetic state and $\sim$1.32\,eV for the ferrimagnetic and conical magnetic states. This emphasizes the important roles played by the magnetic spins for the insulating state of the system.

Hybridization between Fe/Cr-3$d$ and O-2$p$ electrons can be seen due to the bonding in the tetrahedron (FeO$_4$) and octahedron (CrO$_6$) environment. Interestingly, however, the Fe-3$d$ and Cr-3$d$ electrons are located at different energy bands suggesting that they are nearly decoupled.
This is consistent with the fact that the spin-gap transition observed in both our static and dynamic magnetic measurements, due to the breathing Cr$_4$ lattice \cite{prb.96.214439,prl.110.097203}, was not observed in our M\"ossbauer measurements (see the temperature dependence of the hyperfine parameters shown in Fig.\ref{fig:IS} and Fig.\ref{fig:Bhf}). Since the FeO$_4$ and CrO$_6$ units are decoupled and the M\"ossbauer effect only probes the local characteristics at the Fe site, the absence of any anomaly in the hyperfine parameters near the spin-gap transition may be understood naturally.

Furthermore, there might be some anomaly in the M\"ossbauer spectrum at the spin-gap transition if the magnetic structure of the Cr spins changes, which also affects the magnetic structure of the Fe spins through the finite exchange interactions between the Fe and Cr spins. This is true for the ferrimagnetic and magnetostructural transitions where the long-range magnetic structure of Fe/Cr spins changes.
Therefore, the absence of any anomaly in our M\"ossbauer spectrum probably indicates that the long-range magnetic structure of the Cr sublattice  does not change at the spin-gap transition.

For the hyperfine magnetic splitting, the largest contribution to the hyperfine field is the contact term which is proportional to the magnetization density at the nucleus. This is done in the ELK code by directly solving the spin-polarised Dirac equation \cite{prb.35.3271}. When SOC is considered, as in our conical magnetic state calculations, the spin and orbital dipole contributions are added self-consistently to the Kohn-Sham field during the ground state calculation. Then, the hyperfine field $B_{tot}$ can be decomposed theoretically into three parts \cite{prb.35.3271}
\begin{equation}
\label{eqBtot}
B_{tot} = B_{c} + B_{dip} +B_{orb}
\end{equation}
where, $B_c$ is the Fermi contact term, $B_{dip}$ is the spin-dipolar interaction term, and $B_{orb}$ is the spin-orbit correction term.
For Fe$^{3+}$ oxides, the later two parts are usually small \cite{prb.81.174412} and therefore are not reported in this work.

\begin{table}[ht]
\centering
\caption{The calculated magnetic moments $\mu_{cal}$ of the Fe atom and the Fermi contact magnetic fields $B_{c}$ (Tesla) at the Fe site. The experimental values of the Fe magnetic moments are $\mu_{exp}$(Fe)=2.54\,$\mu_B$ and $\mu_{exp}$(Fe)=4.2\,$\mu_B$ for the ferrimagnetic and conical magnetic states, respectively. The experimental values of the hyperfine fields at the Fe site are $B_{exp}=24.3(3)$\,$T$ and $B_{exp}=45.6(3)$\,$T$ for the ferrimagnetic and conical magnetic states, respectively.}
\label{table} {\small
\begin{tabular}{c c c }
\hline \hline
Magnetic states      &  $\mu_{cal}$ ($\mu_B$)   & $B_{c}$ (T)      \\
\hline
Ferromagnetic/DFT+U      &  3.52/4.13          &  26.05/31.56     \\
Ferrimagnetic/DFT+U      &  3.48/3.98          &  47.43/51.50     \\
Conical magnetic/DFT+U   &  3.55/4.04          &  38.42/42.59      \\
\hline \hline
\end{tabular}}
\end{table}

The calculated magnetic moments $\mu_{cal}$ of the Fe atom and the Fermi contact magnetic fields $B_{c}$ at the Fe site are shown in table \ref{table} for the three calculated magnetic structures. For the low temperature conical magnetic state, the calculated magnetic moment and Fermi contact field are close to the corresponding experimental values.
On the other hand, the calculated magnetic moments for the ferro-/ferri- magnetic states are much larger than the experimental value obtained in the ferrimagnetic state. The obtained contact field is also much higher than the experimental value for the ferrimagnetic state. Although the contact field obtained for the ferromagnetic state is close to the experimental value, we believe that this is due to the metallic ground state of the electronic structure for the ferromagnetic state without the U parameter (see Fig.\ref{fig:dos} (c)).
However, it is generally found that the hyperfine magnetic fields at the $^{57}$Fe site roughly scale with its magnetic moments \cite{prb.81.174412}. If we assume the same proportional constant between the hyperfine magnetic field and the magnetic moment for both the ferrimagnetic and conical magnetic states, we arrive at a magnetic moment of $\mu=2.24$\,$\mu_B$ from our M\"ossbauer data. This is close to the value determined by neutron diffractions at 30\,K $\mu_{exp}$(Fe)=2.54\,$\mu_B$ \cite{prb.96.214439}, where the small difference may suggest a slightly different proportional constant.
Anyway, these results indicate that, upon lowering of temperature through $T_{MS}$, there is a spin-state transition for the Fe spins. Since higher spin state of the Fe$^{3+}$ ion has a larger volume, this picture explains naturally why the volume of the local FeO$_4$ tetrahedron is a little larger for the low temperature tetragonal phase \cite{prb.96.214439} than the high temperature cubic phase, whereas a large volume contraction happens when entering the low temperature tetragonal phase from the high temperature cubic phase \cite{jpsj.91.023710}.

\section{Summary}
In summary, we have studied the complex magnetostructural and spin-state transitions of the \lfco\ compound by the combination of magnetization, M\"ossbauer spectroscopy, and DFT calculations.
We observe three magnetic related transitions, namely the ferrimagnetic transition at $T_N\sim94$\,K, the spin-gap transition at $T_{SG}\sim50$\,K, and the magnetostructural transition at $T_{MS}\sim19$\,K from our static $\chi$(T) curve. However, only the first and third transitions were seen from our M\"ossbauer measurements, suggesting that the spin-gap transition is absent at the Fe site.
These results suggest that the spin-gap transition is only an effect of the breathing Cr$_4$ lattice. This is in agreement with our DFT calculations where we see nearly decoupled electronic states for the FeO$_4$ and CrO$_6$ units in all three considered magnetic solutions.
The temperature dependence of the hyperfine magnetic field shows a jump at $T_{MS}$, consistent with a spin-state transition for the Fe spins, which is in agreement with earlier neutron diffraction measurements.

\section{Acknowledgement}
%\textit{Acknowledgement:}
This paper was supported by the National Natural Science Foundation of China (Grants Nos. 11704167, 51971221) and the National Key Research and Development Program of China (Grants No. 2022YFB4101401, 2022YFA1402704).
The authors are grateful to the support provided by the Supercomputing Center of Lanzhou University.

\bibliography{lfco}

\begin{thebibliography}{38}%
\makeatletter
\providecommand \@ifxundefined [1]{%
 \@ifx{#1\undefined}
}%
\providecommand \@ifnum [1]{%
 \ifnum #1\expandafter \@firstoftwo
 \else \expandafter \@secondoftwo
 \fi
}%
\providecommand \@ifx [1]{%
 \ifx #1\expandafter \@firstoftwo
 \else \expandafter \@secondoftwo
 \fi
}%
\providecommand \natexlab [1]{#1}%
\providecommand \enquote  [1]{``#1''}%
\providecommand \bibnamefont  [1]{#1}%
\providecommand \bibfnamefont [1]{#1}%
\providecommand \citenamefont [1]{#1}%
\providecommand \href@noop [0]{\@secondoftwo}%
\providecommand \href [0]{\begingroup \@sanitize@url \@href}%
\providecommand \@href[1]{\@@startlink{#1}\@@href}%
\providecommand \@@href[1]{\endgroup#1\@@endlink}%
\providecommand \@sanitize@url [0]{\catcode `\\12\catcode `\$12\catcode
  `\&12\catcode `\#12\catcode `\^12\catcode `\_12\catcode `\%12\relax}%
\providecommand \@@startlink[1]{}%
\providecommand \@@endlink[0]{}%
\providecommand \url  [0]{\begingroup\@sanitize@url \@url }%
\providecommand \@url [1]{\endgroup\@href {#1}{\urlprefix }}%
\providecommand \urlprefix  [0]{URL }%
\providecommand \Eprint [0]{\href }%
\providecommand \doibase [0]{http://dx.doi.org/}%
\providecommand \selectlanguage [0]{\@gobble}%
\providecommand \bibinfo  [0]{\@secondoftwo}%
\providecommand \bibfield  [0]{\@secondoftwo}%
\providecommand \translation [1]{[#1]}%
\providecommand \BibitemOpen [0]{}%
\providecommand \bibitemStop [0]{}%
\providecommand \bibitemNoStop [0]{.\EOS\space}%
\providecommand \EOS [0]{\spacefactor3000\relax}%
\providecommand \BibitemShut  [1]{\csname bibitem#1\endcsname}%
\let\auto@bib@innerbib\@empty
%</preamble>
\bibitem [{\citenamefont {Gardner}\ \emph {et~al.}(2010)\citenamefont
  {Gardner}, \citenamefont {Gingras},\ and\ \citenamefont
  {Greedan}}]{rmp.82.53}%
  \BibitemOpen
  \bibfield  {author} {\bibinfo {author} {\bibfnamefont {J.~S.}\ \bibnamefont
  {Gardner}}, \bibinfo {author} {\bibfnamefont {M.~J.~P.}\ \bibnamefont
  {Gingras}}, \ and\ \bibinfo {author} {\bibfnamefont {J.~E.}\ \bibnamefont
  {Greedan}},\ }\href {\doibase 10.1103/RevModPhys.82.53} {\bibfield  {journal}
  {\bibinfo  {journal} {Rev. Mod. Phys.}\ }\textbf {\bibinfo {volume} {82}},\
  \bibinfo {pages} {53} (\bibinfo {year} {2010})}\BibitemShut {NoStop}%
\bibitem [{\citenamefont {Lee}\ \emph {et~al.}(2010)\citenamefont {Lee},
  \citenamefont {Takagi}, \citenamefont {Louca}, \citenamefont {Matsuda},
  \citenamefont {Ji}, \citenamefont {Ueda}, \citenamefont {Ueda}, \citenamefont
  {Katsufuji}, \citenamefont {Chung}, \citenamefont {Park}, \citenamefont
  {Cheong},\ and\ \citenamefont {Broholm}}]{jpsj.79.011004}%
  \BibitemOpen
  \bibfield  {author} {\bibinfo {author} {\bibfnamefont {S.-H.}\ \bibnamefont
  {Lee}}, \bibinfo {author} {\bibfnamefont {H.}~\bibnamefont {Takagi}},
  \bibinfo {author} {\bibfnamefont {D.}~\bibnamefont {Louca}}, \bibinfo
  {author} {\bibfnamefont {M.}~\bibnamefont {Matsuda}}, \bibinfo {author}
  {\bibfnamefont {S.}~\bibnamefont {Ji}}, \bibinfo {author} {\bibfnamefont
  {H.}~\bibnamefont {Ueda}}, \bibinfo {author} {\bibfnamefont {Y.}~\bibnamefont
  {Ueda}}, \bibinfo {author} {\bibfnamefont {T.}~\bibnamefont {Katsufuji}},
  \bibinfo {author} {\bibfnamefont {J.-H.}\ \bibnamefont {Chung}}, \bibinfo
  {author} {\bibfnamefont {S.}~\bibnamefont {Park}}, \bibinfo {author}
  {\bibfnamefont {S.-W.}\ \bibnamefont {Cheong}}, \ and\ \bibinfo {author}
  {\bibfnamefont {C.}~\bibnamefont {Broholm}},\ }\href {\doibase
  10.1143/JPSJ.79.011004} {\bibfield  {journal} {\bibinfo  {journal} {Journal
  of the Physical Society of Japan}\ }\textbf {\bibinfo {volume} {79}},\
  \bibinfo {pages} {011004} (\bibinfo {year} {2010})},\ \Eprint
  {http://arxiv.org/abs/https://doi.org/10.1143/JPSJ.79.011004}
  {https://doi.org/10.1143/JPSJ.79.011004} \BibitemShut {NoStop}%
\bibitem [{\citenamefont {Okamoto}\ \emph {et~al.}(2013)\citenamefont
  {Okamoto}, \citenamefont {Nilsen}, \citenamefont {Attfield},\ and\
  \citenamefont {Hiroi}}]{prl.110.097203}%
  \BibitemOpen
  \bibfield  {author} {\bibinfo {author} {\bibfnamefont {Y.}~\bibnamefont
  {Okamoto}}, \bibinfo {author} {\bibfnamefont {G.~J.}\ \bibnamefont {Nilsen}},
  \bibinfo {author} {\bibfnamefont {J.~P.}\ \bibnamefont {Attfield}}, \ and\
  \bibinfo {author} {\bibfnamefont {Z.}~\bibnamefont {Hiroi}},\ }\href
  {\doibase 10.1103/PhysRevLett.110.097203} {\bibfield  {journal} {\bibinfo
  {journal} {Phys. Rev. Lett.}\ }\textbf {\bibinfo {volume} {110}},\ \bibinfo
  {pages} {097203} (\bibinfo {year} {2013})}\BibitemShut {NoStop}%
\bibitem [{\citenamefont {Sharma}\ \emph {et~al.}(2022)\citenamefont {Sharma},
  \citenamefont {Pocrnic}, \citenamefont {Richtik}, \citenamefont {Wiebe},
  \citenamefont {Beare}, \citenamefont {Gautreau}, \citenamefont {Clancy},
  \citenamefont {Ruff}, \citenamefont {Pula}, \citenamefont {Chen},
  \citenamefont {Yoon}, \citenamefont {Cai},\ and\ \citenamefont
  {Luke}}]{prb.106.024407}%
  \BibitemOpen
  \bibfield  {author} {\bibinfo {author} {\bibfnamefont {S.}~\bibnamefont
  {Sharma}}, \bibinfo {author} {\bibfnamefont {M.}~\bibnamefont {Pocrnic}},
  \bibinfo {author} {\bibfnamefont {B.~N.}\ \bibnamefont {Richtik}}, \bibinfo
  {author} {\bibfnamefont {C.~R.}\ \bibnamefont {Wiebe}}, \bibinfo {author}
  {\bibfnamefont {J.}~\bibnamefont {Beare}}, \bibinfo {author} {\bibfnamefont
  {J.}~\bibnamefont {Gautreau}}, \bibinfo {author} {\bibfnamefont {J.~P.}\
  \bibnamefont {Clancy}}, \bibinfo {author} {\bibfnamefont {J.~P.~C.}\
  \bibnamefont {Ruff}}, \bibinfo {author} {\bibfnamefont {M.}~\bibnamefont
  {Pula}}, \bibinfo {author} {\bibfnamefont {Q.}~\bibnamefont {Chen}}, \bibinfo
  {author} {\bibfnamefont {S.}~\bibnamefont {Yoon}}, \bibinfo {author}
  {\bibfnamefont {Y.}~\bibnamefont {Cai}}, \ and\ \bibinfo {author}
  {\bibfnamefont {G.~M.}\ \bibnamefont {Luke}},\ }\href {\doibase
  10.1103/PhysRevB.106.024407} {\bibfield  {journal} {\bibinfo  {journal}
  {Phys. Rev. B}\ }\textbf {\bibinfo {volume} {106}},\ \bibinfo {pages}
  {024407} (\bibinfo {year} {2022})}\BibitemShut {NoStop}%
\bibitem [{\citenamefont {Aoyama}\ and\ \citenamefont
  {Kawamura}(2021)}]{prb.103.014406}%
  \BibitemOpen
  \bibfield  {author} {\bibinfo {author} {\bibfnamefont {K.}~\bibnamefont
  {Aoyama}}\ and\ \bibinfo {author} {\bibfnamefont {H.}~\bibnamefont
  {Kawamura}},\ }\href {\doibase 10.1103/PhysRevB.103.014406} {\bibfield
  {journal} {\bibinfo  {journal} {Phys. Rev. B}\ }\textbf {\bibinfo {volume}
  {103}},\ \bibinfo {pages} {014406} (\bibinfo {year} {2021})}\BibitemShut
  {NoStop}%
\bibitem [{\citenamefont {Aoyama}\ and\ \citenamefont
  {Kawamura}(2022)}]{prb.106.064412}%
  \BibitemOpen
  \bibfield  {author} {\bibinfo {author} {\bibfnamefont {K.}~\bibnamefont
  {Aoyama}}\ and\ \bibinfo {author} {\bibfnamefont {H.}~\bibnamefont
  {Kawamura}},\ }\href {\doibase 10.1103/PhysRevB.106.064412} {\bibfield
  {journal} {\bibinfo  {journal} {Phys. Rev. B}\ }\textbf {\bibinfo {volume}
  {106}},\ \bibinfo {pages} {064412} (\bibinfo {year} {2022})}\BibitemShut
  {NoStop}%
\bibitem [{\citenamefont {He}\ \emph {et~al.}(2021)\citenamefont {He},
  \citenamefont {Gu}, \citenamefont {Wo}, \citenamefont {Feng}, \citenamefont
  {Hu}, \citenamefont {Hao}, \citenamefont {Gu}, \citenamefont {Walker},
  \citenamefont {Adroja},\ and\ \citenamefont {Zhao}}]{prl.127.147205}%
  \BibitemOpen
  \bibfield  {author} {\bibinfo {author} {\bibfnamefont {Z.}~\bibnamefont
  {He}}, \bibinfo {author} {\bibfnamefont {Y.}~\bibnamefont {Gu}}, \bibinfo
  {author} {\bibfnamefont {H.}~\bibnamefont {Wo}}, \bibinfo {author}
  {\bibfnamefont {Y.}~\bibnamefont {Feng}}, \bibinfo {author} {\bibfnamefont
  {D.}~\bibnamefont {Hu}}, \bibinfo {author} {\bibfnamefont {Y.}~\bibnamefont
  {Hao}}, \bibinfo {author} {\bibfnamefont {Y.}~\bibnamefont {Gu}}, \bibinfo
  {author} {\bibfnamefont {H.~C.}\ \bibnamefont {Walker}}, \bibinfo {author}
  {\bibfnamefont {D.~T.}\ \bibnamefont {Adroja}}, \ and\ \bibinfo {author}
  {\bibfnamefont {J.}~\bibnamefont {Zhao}},\ }\href {\doibase
  10.1103/PhysRevLett.127.147205} {\bibfield  {journal} {\bibinfo  {journal}
  {Phys. Rev. Lett.}\ }\textbf {\bibinfo {volume} {127}},\ \bibinfo {pages}
  {147205} (\bibinfo {year} {2021})}\BibitemShut {NoStop}%
\bibitem [{\citenamefont {Okamoto}\ \emph {et~al.}(2015)\citenamefont
  {Okamoto}, \citenamefont {Nilsen}, \citenamefont {Nakazono},\ and\
  \citenamefont {Hiroi}}]{jpsj.84.043707}%
  \BibitemOpen
  \bibfield  {author} {\bibinfo {author} {\bibfnamefont {Y.}~\bibnamefont
  {Okamoto}}, \bibinfo {author} {\bibfnamefont {G.~J.}\ \bibnamefont {Nilsen}},
  \bibinfo {author} {\bibfnamefont {T.}~\bibnamefont {Nakazono}}, \ and\
  \bibinfo {author} {\bibfnamefont {Z.}~\bibnamefont {Hiroi}},\ }\href
  {\doibase 10.7566/JPSJ.84.043707} {\bibfield  {journal} {\bibinfo  {journal}
  {Journal of the Physical Society of Japan}\ }\textbf {\bibinfo {volume}
  {84}},\ \bibinfo {pages} {043707} (\bibinfo {year} {2015})},\ \Eprint
  {http://arxiv.org/abs/https://doi.org/10.7566/JPSJ.84.043707}
  {https://doi.org/10.7566/JPSJ.84.043707} \BibitemShut {NoStop}%
\bibitem [{\citenamefont {Yaresko}(2008)}]{prb.77.115106}%
  \BibitemOpen
  \bibfield  {author} {\bibinfo {author} {\bibfnamefont {A.~N.}\ \bibnamefont
  {Yaresko}},\ }\href {\doibase 10.1103/PhysRevB.77.115106} {\bibfield
  {journal} {\bibinfo  {journal} {Phys. Rev. B}\ }\textbf {\bibinfo {volume}
  {77}},\ \bibinfo {pages} {115106} (\bibinfo {year} {2008})}\BibitemShut
  {NoStop}%
\bibitem [{\citenamefont {Pokharel}\ \emph {et~al.}(2020)\citenamefont
  {Pokharel}, \citenamefont {Arachchige}, \citenamefont {Williams},
  \citenamefont {May}, \citenamefont {Fishman}, \citenamefont {Sala},
  \citenamefont {Calder}, \citenamefont {Ehlers}, \citenamefont {Parker},
  \citenamefont {Hong}, \citenamefont {Wildes}, \citenamefont {Mandrus},
  \citenamefont {Paddison},\ and\ \citenamefont
  {Christianson}}]{prl.125.167201}%
  \BibitemOpen
  \bibfield  {author} {\bibinfo {author} {\bibfnamefont {G.}~\bibnamefont
  {Pokharel}}, \bibinfo {author} {\bibfnamefont {H.~S.}\ \bibnamefont
  {Arachchige}}, \bibinfo {author} {\bibfnamefont {T.~J.}\ \bibnamefont
  {Williams}}, \bibinfo {author} {\bibfnamefont {A.~F.}\ \bibnamefont {May}},
  \bibinfo {author} {\bibfnamefont {R.~S.}\ \bibnamefont {Fishman}}, \bibinfo
  {author} {\bibfnamefont {G.}~\bibnamefont {Sala}}, \bibinfo {author}
  {\bibfnamefont {S.}~\bibnamefont {Calder}}, \bibinfo {author} {\bibfnamefont
  {G.}~\bibnamefont {Ehlers}}, \bibinfo {author} {\bibfnamefont {D.~S.}\
  \bibnamefont {Parker}}, \bibinfo {author} {\bibfnamefont {T.}~\bibnamefont
  {Hong}}, \bibinfo {author} {\bibfnamefont {A.}~\bibnamefont {Wildes}},
  \bibinfo {author} {\bibfnamefont {D.}~\bibnamefont {Mandrus}}, \bibinfo
  {author} {\bibfnamefont {J.~A.~M.}\ \bibnamefont {Paddison}}, \ and\ \bibinfo
  {author} {\bibfnamefont {A.~D.}\ \bibnamefont {Christianson}},\ }\href
  {\doibase 10.1103/PhysRevLett.125.167201} {\bibfield  {journal} {\bibinfo
  {journal} {Phys. Rev. Lett.}\ }\textbf {\bibinfo {volume} {125}},\ \bibinfo
  {pages} {167201} (\bibinfo {year} {2020})}\BibitemShut {NoStop}%
\bibitem [{\citenamefont {Ghosh}\ \emph {et~al.}(2019)\citenamefont {Ghosh},
  \citenamefont {Iqbal}, \citenamefont {M\"uller}, \citenamefont {Ponnaganti},
  \citenamefont {Thomale}, \citenamefont {Narayanan}, \citenamefont {Reuther},
  \citenamefont {Gingras},\ and\ \citenamefont {Jeschke}}]{npj.4.63}%
  \BibitemOpen
  \bibfield  {author} {\bibinfo {author} {\bibfnamefont {P.}~\bibnamefont
  {Ghosh}}, \bibinfo {author} {\bibfnamefont {Y.}~\bibnamefont {Iqbal}},
  \bibinfo {author} {\bibfnamefont {T.}~\bibnamefont {M\"uller}}, \bibinfo
  {author} {\bibfnamefont {R.~T.}\ \bibnamefont {Ponnaganti}}, \bibinfo
  {author} {\bibfnamefont {R.}~\bibnamefont {Thomale}}, \bibinfo {author}
  {\bibfnamefont {R.}~\bibnamefont {Narayanan}}, \bibinfo {author}
  {\bibfnamefont {J.}~\bibnamefont {Reuther}}, \bibinfo {author} {\bibfnamefont
  {M.~J.~P.}\ \bibnamefont {Gingras}}, \ and\ \bibinfo {author} {\bibfnamefont
  {H.~O.}\ \bibnamefont {Jeschke}},\ }\href {\doibase
  10.1038/s41535-019-0202-z} {\bibfield  {journal} {\bibinfo  {journal} {npj
  Quantum Materials}\ }\textbf {\bibinfo {volume} {4}},\ \bibinfo {pages} {63}
  (\bibinfo {year} {2019})}\BibitemShut {NoStop}%
\bibitem [{\citenamefont {Feng}\ \emph {et~al.}(2020)\citenamefont {Feng},
  \citenamefont {Liu}, \citenamefont {Bian}, \citenamefont {Xiong},
  \citenamefont {Zhu}, \citenamefont {Zong}, \citenamefont {Shi},\ and\
  \citenamefont {Fang}}]{pssb.257.1900685}%
  \BibitemOpen
  \bibfield  {author} {\bibinfo {author} {\bibfnamefont {Y.}~\bibnamefont
  {Feng}}, \bibinfo {author} {\bibfnamefont {H.}~\bibnamefont {Liu}}, \bibinfo
  {author} {\bibfnamefont {J.}~\bibnamefont {Bian}}, \bibinfo {author}
  {\bibfnamefont {W.}~\bibnamefont {Xiong}}, \bibinfo {author} {\bibfnamefont
  {S.}~\bibnamefont {Zhu}}, \bibinfo {author} {\bibfnamefont {B.}~\bibnamefont
  {Zong}}, \bibinfo {author} {\bibfnamefont {B.}~\bibnamefont {Shi}}, \ and\
  \bibinfo {author} {\bibfnamefont {B.}~\bibnamefont {Fang}},\ }\href {\doibase
  https://doi.org/10.1002/pssb.201900685} {\bibfield  {journal} {\bibinfo
  {journal} {physica status solidi (b)}\ }\textbf {\bibinfo {volume} {257}},\
  \bibinfo {pages} {1900685} (\bibinfo {year} {2020})},\ \Eprint
  {http://arxiv.org/abs/https://onlinelibrary.wiley.com/doi/pdf/10.1002/pssb.201900685}
  {https://onlinelibrary.wiley.com/doi/pdf/10.1002/pssb.201900685} \BibitemShut
  {NoStop}%
\bibitem [{\citenamefont {Feng}\ \emph {et~al.}(2022)\citenamefont {Feng},
  \citenamefont {Zuo}, \citenamefont {Bian}, \citenamefont {Zhang},
  \citenamefont {Li}, \citenamefont {Huang}, \citenamefont {Fang},\ and\
  \citenamefont {Liu}}]{rp.35.105379}%
  \BibitemOpen
  \bibfield  {author} {\bibinfo {author} {\bibfnamefont {Y.}~\bibnamefont
  {Feng}}, \bibinfo {author} {\bibfnamefont {Z.}~\bibnamefont {Zuo}}, \bibinfo
  {author} {\bibfnamefont {J.}~\bibnamefont {Bian}}, \bibinfo {author}
  {\bibfnamefont {L.}~\bibnamefont {Zhang}}, \bibinfo {author} {\bibfnamefont
  {J.}~\bibnamefont {Li}}, \bibinfo {author} {\bibfnamefont {Y.}~\bibnamefont
  {Huang}}, \bibinfo {author} {\bibfnamefont {B.}~\bibnamefont {Fang}}, \ and\
  \bibinfo {author} {\bibfnamefont {H.}~\bibnamefont {Liu}},\ }\href {\doibase
  https://doi.org/10.1016/j.rinp.2022.105379} {\bibfield  {journal} {\bibinfo
  {journal} {Results in Physics}\ }\textbf {\bibinfo {volume} {35}},\ \bibinfo
  {pages} {105379} (\bibinfo {year} {2022})}\BibitemShut {NoStop}%
\bibitem [{\citenamefont {Wang}\ \emph {et~al.}(2016)\citenamefont {Wang},
  \citenamefont {Tan}, \citenamefont {Huang},\ and\ \citenamefont
  {Shu}}]{cpl.33.127501}%
  \BibitemOpen
  \bibfield  {author} {\bibinfo {author} {\bibfnamefont {D.-Y.}\ \bibnamefont
  {Wang}}, \bibinfo {author} {\bibfnamefont {C.}~\bibnamefont {Tan}}, \bibinfo
  {author} {\bibfnamefont {K.}~\bibnamefont {Huang}}, \ and\ \bibinfo {author}
  {\bibfnamefont {L.}~\bibnamefont {Shu}},\ }\href {\doibase
  10.1088/0256-307X/33/12/127501} {\bibfield  {journal} {\bibinfo  {journal}
  {Chinese Physics Letters}\ }\textbf {\bibinfo {volume} {33}},\ \bibinfo
  {pages} {127501} (\bibinfo {year} {2016})}\BibitemShut {NoStop}%
\bibitem [{\citenamefont {Gen}\ \emph {et~al.}(2020)\citenamefont {Gen},
  \citenamefont {Okamoto}, \citenamefont {Mori}, \citenamefont {Takenaka},\
  and\ \citenamefont {Kohama}}]{prb.101.054434}%
  \BibitemOpen
  \bibfield  {author} {\bibinfo {author} {\bibfnamefont {M.}~\bibnamefont
  {Gen}}, \bibinfo {author} {\bibfnamefont {Y.}~\bibnamefont {Okamoto}},
  \bibinfo {author} {\bibfnamefont {M.}~\bibnamefont {Mori}}, \bibinfo {author}
  {\bibfnamefont {K.}~\bibnamefont {Takenaka}}, \ and\ \bibinfo {author}
  {\bibfnamefont {Y.}~\bibnamefont {Kohama}},\ }\href {\doibase
  10.1103/PhysRevB.101.054434} {\bibfield  {journal} {\bibinfo  {journal}
  {Phys. Rev. B}\ }\textbf {\bibinfo {volume} {101}},\ \bibinfo {pages}
  {054434} (\bibinfo {year} {2020})}\BibitemShut {NoStop}%
\bibitem [{\citenamefont {Okamoto}\ \emph {et~al.}(2022)\citenamefont
  {Okamoto}, \citenamefont {Kanematsu}, \citenamefont {Kubota}, \citenamefont
  {Yajima},\ and\ \citenamefont {Takenaka}}]{jpsj.91.023710}%
  \BibitemOpen
  \bibfield  {author} {\bibinfo {author} {\bibfnamefont {Y.}~\bibnamefont
  {Okamoto}}, \bibinfo {author} {\bibfnamefont {T.}~\bibnamefont {Kanematsu}},
  \bibinfo {author} {\bibfnamefont {Y.}~\bibnamefont {Kubota}}, \bibinfo
  {author} {\bibfnamefont {T.}~\bibnamefont {Yajima}}, \ and\ \bibinfo {author}
  {\bibfnamefont {K.}~\bibnamefont {Takenaka}},\ }\href {\doibase
  10.7566/JPSJ.91.023710} {\bibfield  {journal} {\bibinfo  {journal} {Journal
  of the Physical Society of Japan}\ }\textbf {\bibinfo {volume} {91}},\
  \bibinfo {pages} {023710} (\bibinfo {year} {2022})},\ \Eprint
  {http://arxiv.org/abs/https://doi.org/10.7566/JPSJ.91.023710}
  {https://doi.org/10.7566/JPSJ.91.023710} \BibitemShut {NoStop}%
\bibitem [{\citenamefont {Saha}\ \emph {et~al.}(2017)\citenamefont {Saha},
  \citenamefont {Dhanya}, \citenamefont {Bellin}, \citenamefont {B\'eneut},
  \citenamefont {Bhattacharyya}, \citenamefont {Shukla}, \citenamefont
  {Narayana}, \citenamefont {Suard}, \citenamefont {Rodr\'{\i}guez-Carvajal},\
  and\ \citenamefont {Sundaresan}}]{prb.96.214439}%
  \BibitemOpen
  \bibfield  {author} {\bibinfo {author} {\bibfnamefont {R.}~\bibnamefont
  {Saha}}, \bibinfo {author} {\bibfnamefont {R.}~\bibnamefont {Dhanya}},
  \bibinfo {author} {\bibfnamefont {C.}~\bibnamefont {Bellin}}, \bibinfo
  {author} {\bibfnamefont {K.}~\bibnamefont {B\'eneut}}, \bibinfo {author}
  {\bibfnamefont {A.}~\bibnamefont {Bhattacharyya}}, \bibinfo {author}
  {\bibfnamefont {A.}~\bibnamefont {Shukla}}, \bibinfo {author} {\bibfnamefont
  {C.}~\bibnamefont {Narayana}}, \bibinfo {author} {\bibfnamefont
  {E.}~\bibnamefont {Suard}}, \bibinfo {author} {\bibfnamefont
  {J.}~\bibnamefont {Rodr\'{\i}guez-Carvajal}}, \ and\ \bibinfo {author}
  {\bibfnamefont {A.}~\bibnamefont {Sundaresan}},\ }\href {\doibase
  10.1103/PhysRevB.96.214439} {\bibfield  {journal} {\bibinfo  {journal} {Phys.
  Rev. B}\ }\textbf {\bibinfo {volume} {96}},\ \bibinfo {pages} {214439}
  (\bibinfo {year} {2017})}\BibitemShut {NoStop}%
\bibitem [{\citenamefont {Momma}\ and\ \citenamefont {Izumi}(2011)}]{vesta}%
  \BibitemOpen
  \bibfield  {author} {\bibinfo {author} {\bibfnamefont {K.}~\bibnamefont
  {Momma}}\ and\ \bibinfo {author} {\bibfnamefont {F.}~\bibnamefont {Izumi}},\
  }\href {\doibase 10.1107/S0021889811038970} {\bibfield  {journal} {\bibinfo
  {journal} {Journal of Applied Crystallography}\ }\textbf {\bibinfo {volume}
  {44}},\ \bibinfo {pages} {1272} (\bibinfo {year} {2011})}\BibitemShut
  {NoStop}%
\bibitem [{ful()}]{fullprof}%
  \BibitemOpen
  \href@noop {} {\enquote {\bibinfo {title} {{The FullProf Suite}},}\ }\bibinfo
  {howpublished} {\url{http://www.ill.eu/sites/fullprof/}}\BibitemShut
  {NoStop}%
\bibitem [{elk()}]{elk}%
  \BibitemOpen
  \href@noop {} {\enquote {\bibinfo {title} {{The Elk Code}},}\ }\bibinfo
  {howpublished} {\url{http://elk.sourceforge.net/}}\BibitemShut {NoStop}%
\bibitem [{\citenamefont {Perdew}\ and\ \citenamefont
  {Wang}(1992)}]{prb.45.13244}%
  \BibitemOpen
  \bibfield  {author} {\bibinfo {author} {\bibfnamefont {J.~P.}\ \bibnamefont
  {Perdew}}\ and\ \bibinfo {author} {\bibfnamefont {Y.}~\bibnamefont {Wang}},\
  }\href {\doibase 10.1103/PhysRevB.45.13244} {\bibfield  {journal} {\bibinfo
  {journal} {Phys. Rev. B}\ }\textbf {\bibinfo {volume} {45}},\ \bibinfo
  {pages} {13244} (\bibinfo {year} {1992})}\BibitemShut {NoStop}%
\bibitem [{\citenamefont {Liechtenstein}\ \emph {et~al.}(1995)\citenamefont
  {Liechtenstein}, \citenamefont {Anisimov},\ and\ \citenamefont
  {Zaanen}}]{prb.52.R5467}%
  \BibitemOpen
  \bibfield  {author} {\bibinfo {author} {\bibfnamefont {A.~I.}\ \bibnamefont
  {Liechtenstein}}, \bibinfo {author} {\bibfnamefont {V.~I.}\ \bibnamefont
  {Anisimov}}, \ and\ \bibinfo {author} {\bibfnamefont {J.}~\bibnamefont
  {Zaanen}},\ }\href {\doibase 10.1103/PhysRevB.52.R5467} {\bibfield  {journal}
  {\bibinfo  {journal} {Phys. Rev. B}\ }\textbf {\bibinfo {volume} {52}},\
  \bibinfo {pages} {R5467} (\bibinfo {year} {1995})}\BibitemShut {NoStop}%
\bibitem [{\citenamefont {{Marinkovi\'{c} Stanojevi\'{c} }}\ \emph
  {et~al.}(2007)\citenamefont {{Marinkovi\'{c} Stanojevi\'{c} }}, \citenamefont
  {Rom\v{c}evi\'{c}},\ and\ \citenamefont {Stojanovi\'{c}}}]{jecs.27.903}%
  \BibitemOpen
  \bibfield  {author} {\bibinfo {author} {\bibfnamefont {Z.}~\bibnamefont
  {{Marinkovi\'{c} Stanojevi\'{c} }}}, \bibinfo {author} {\bibfnamefont
  {N.}~\bibnamefont {Rom\v{c}evi\'{c}}}, \ and\ \bibinfo {author}
  {\bibfnamefont {B.}~\bibnamefont {Stojanovi\'{c}}},\ }\href {\doibase
  https://doi.org/10.1016/j.jeurceramsoc.2006.04.057} {\bibfield  {journal}
  {\bibinfo  {journal} {Journal of the European Ceramic Society}\ }\textbf
  {\bibinfo {volume} {27}},\ \bibinfo {pages} {903} (\bibinfo {year} {2007})},\
  \bibinfo {note} {refereed Reports IX Conference \& Exhibition of the European
  Ceramic Society}\BibitemShut {NoStop}%
\bibitem [{\citenamefont {Kumar}\ and\ \citenamefont
  {Yusuf}(2015)}]{prep.556.1}%
  \BibitemOpen
  \bibfield  {author} {\bibinfo {author} {\bibfnamefont {A.}~\bibnamefont
  {Kumar}}\ and\ \bibinfo {author} {\bibfnamefont {S.}~\bibnamefont {Yusuf}},\
  }\href {\doibase https://doi.org/10.1016/j.physrep.2014.10.003} {\bibfield
  {journal} {\bibinfo  {journal} {Physics Reports}\ }\textbf {\bibinfo {volume}
  {556}},\ \bibinfo {pages} {1} (\bibinfo {year} {2015})},\ \bibinfo {note}
  {the phenomenon of negative magnetization and its implications}\BibitemShut
  {NoStop}%
\bibitem [{\citenamefont {Saha}\ \emph {et~al.}(2016)\citenamefont {Saha},
  \citenamefont {Fauth}, \citenamefont {Avdeev}, \citenamefont {Kayser},
  \citenamefont {Kennedy},\ and\ \citenamefont {Sundaresan}}]{prb.94.064420}%
  \BibitemOpen
  \bibfield  {author} {\bibinfo {author} {\bibfnamefont {R.}~\bibnamefont
  {Saha}}, \bibinfo {author} {\bibfnamefont {F.}~\bibnamefont {Fauth}},
  \bibinfo {author} {\bibfnamefont {M.}~\bibnamefont {Avdeev}}, \bibinfo
  {author} {\bibfnamefont {P.}~\bibnamefont {Kayser}}, \bibinfo {author}
  {\bibfnamefont {B.~J.}\ \bibnamefont {Kennedy}}, \ and\ \bibinfo {author}
  {\bibfnamefont {A.}~\bibnamefont {Sundaresan}},\ }\href {\doibase
  10.1103/PhysRevB.94.064420} {\bibfield  {journal} {\bibinfo  {journal} {Phys.
  Rev. B}\ }\textbf {\bibinfo {volume} {94}},\ \bibinfo {pages} {064420}
  (\bibinfo {year} {2016})}\BibitemShut {NoStop}%
\bibitem [{\citenamefont {Nilsen}\ \emph {et~al.}(2015)\citenamefont {Nilsen},
  \citenamefont {Okamoto}, \citenamefont {Masuda}, \citenamefont
  {Rodriguez-Carvajal}, \citenamefont {Mutka}, \citenamefont {Hansen},\ and\
  \citenamefont {Hiroi}}]{prb.91.174435}%
  \BibitemOpen
  \bibfield  {author} {\bibinfo {author} {\bibfnamefont {G.~J.}\ \bibnamefont
  {Nilsen}}, \bibinfo {author} {\bibfnamefont {Y.}~\bibnamefont {Okamoto}},
  \bibinfo {author} {\bibfnamefont {T.}~\bibnamefont {Masuda}}, \bibinfo
  {author} {\bibfnamefont {J.}~\bibnamefont {Rodriguez-Carvajal}}, \bibinfo
  {author} {\bibfnamefont {H.}~\bibnamefont {Mutka}}, \bibinfo {author}
  {\bibfnamefont {T.}~\bibnamefont {Hansen}}, \ and\ \bibinfo {author}
  {\bibfnamefont {Z.}~\bibnamefont {Hiroi}},\ }\href {\doibase
  10.1103/PhysRevB.91.174435} {\bibfield  {journal} {\bibinfo  {journal} {Phys.
  Rev. B}\ }\textbf {\bibinfo {volume} {91}},\ \bibinfo {pages} {174435}
  (\bibinfo {year} {2015})}\BibitemShut {NoStop}%
\bibitem [{\citenamefont {Garc\'{\i}a-Palacios}\ and\ \citenamefont
  {L\'azaro}(1998)}]{prb.58.14937}%
  \BibitemOpen
  \bibfield  {author} {\bibinfo {author} {\bibfnamefont {J.~L.}\ \bibnamefont
  {Garc\'{\i}a-Palacios}}\ and\ \bibinfo {author} {\bibfnamefont {F.~J.}\
  \bibnamefont {L\'azaro}},\ }\href {\doibase 10.1103/PhysRevB.58.14937}
  {\bibfield  {journal} {\bibinfo  {journal} {Phys. Rev. B}\ }\textbf {\bibinfo
  {volume} {58}},\ \bibinfo {pages} {14937} (\bibinfo {year}
  {1998})}\BibitemShut {NoStop}%
\bibitem [{\citenamefont {Kunikin}\ \emph {et~al.}(2019)\citenamefont
  {Kunikin}, \citenamefont {Zakinyan},\ and\ \citenamefont
  {Dikansky}}]{jmmm.483.178}%
  \BibitemOpen
  \bibfield  {author} {\bibinfo {author} {\bibfnamefont {S.}~\bibnamefont
  {Kunikin}}, \bibinfo {author} {\bibfnamefont {A.}~\bibnamefont {Zakinyan}}, \
  and\ \bibinfo {author} {\bibfnamefont {Y.}~\bibnamefont {Dikansky}},\ }\href
  {\doibase https://doi.org/10.1016/j.jmmm.2019.03.116} {\bibfield  {journal}
  {\bibinfo  {journal} {Journal of Magnetism and Magnetic Materials}\ }\textbf
  {\bibinfo {volume} {483}},\ \bibinfo {pages} {178} (\bibinfo {year}
  {2019})}\BibitemShut {NoStop}%
\bibitem [{\citenamefont {Sobolev}\ \emph {et~al.}(2018)\citenamefont
  {Sobolev}, \citenamefont {Akulenko}, \citenamefont {Glazkova}, \citenamefont
  {Pankratov},\ and\ \citenamefont {Presniakov}}]{prb.97.104415}%
  \BibitemOpen
  \bibfield  {author} {\bibinfo {author} {\bibfnamefont {A.~V.}\ \bibnamefont
  {Sobolev}}, \bibinfo {author} {\bibfnamefont {A.~A.}\ \bibnamefont
  {Akulenko}}, \bibinfo {author} {\bibfnamefont {I.~S.}\ \bibnamefont
  {Glazkova}}, \bibinfo {author} {\bibfnamefont {D.~A.}\ \bibnamefont
  {Pankratov}}, \ and\ \bibinfo {author} {\bibfnamefont {I.~A.}\ \bibnamefont
  {Presniakov}},\ }\href {\doibase 10.1103/PhysRevB.97.104415} {\bibfield
  {journal} {\bibinfo  {journal} {Phys. Rev. B}\ }\textbf {\bibinfo {volume}
  {97}},\ \bibinfo {pages} {104415} (\bibinfo {year} {2018})}\BibitemShut
  {NoStop}%
\bibitem [{\citenamefont {G\"{u}tlich}\ \emph {et~al.}()\citenamefont
  {G\"{u}tlich}, \citenamefont {Bill},\ and\ \citenamefont
  {Trautwein}}]{PGmbook2011}%
  \BibitemOpen
  \bibfield  {author} {\bibinfo {author} {\bibfnamefont {P.}~\bibnamefont
  {G\"{u}tlich}}, \bibinfo {author} {\bibfnamefont {E.}~\bibnamefont {Bill}}, \
  and\ \bibinfo {author} {\bibfnamefont {A.~X.}\ \bibnamefont {Trautwein}},\
  }\href {\doibase 10.1007/978-3-540-88428-6} {\emph {\bibinfo {title}
  {M\"ossbauer Spectroscopy and Transition Metal Chemistry}}}\ (\bibinfo
  {publisher} {Springer Berlin Heidelberg})\BibitemShut {NoStop}%
\bibitem [{\citenamefont {Le~Guillou}\ and\ \citenamefont
  {Zinn-Justin}(1977)}]{prl.39.95}%
  \BibitemOpen
  \bibfield  {author} {\bibinfo {author} {\bibfnamefont {J.~C.}\ \bibnamefont
  {Le~Guillou}}\ and\ \bibinfo {author} {\bibfnamefont {J.}~\bibnamefont
  {Zinn-Justin}},\ }\href {\doibase 10.1103/PhysRevLett.39.95} {\bibfield
  {journal} {\bibinfo  {journal} {Phys. Rev. Lett.}\ }\textbf {\bibinfo
  {volume} {39}},\ \bibinfo {pages} {95} (\bibinfo {year} {1977})}\BibitemShut
  {NoStop}%
\bibitem [{\citenamefont {Le~Guillou}\ and\ \citenamefont
  {Zinn-Justin}(1980)}]{prb.21.3976}%
  \BibitemOpen
  \bibfield  {author} {\bibinfo {author} {\bibfnamefont {J.~C.}\ \bibnamefont
  {Le~Guillou}}\ and\ \bibinfo {author} {\bibfnamefont {J.}~\bibnamefont
  {Zinn-Justin}},\ }\href {\doibase 10.1103/PhysRevB.21.3976} {\bibfield
  {journal} {\bibinfo  {journal} {Phys. Rev. B}\ }\textbf {\bibinfo {volume}
  {21}},\ \bibinfo {pages} {3976} (\bibinfo {year} {1980})}\BibitemShut
  {NoStop}%
\bibitem [{\citenamefont {Baxter}(1982)}]{jpa.15.3329}%
  \BibitemOpen
  \bibfield  {author} {\bibinfo {author} {\bibfnamefont {R.~J.}\ \bibnamefont
  {Baxter}},\ }\href {\doibase 10.1088/0305-4470/15/10/035} {\bibfield
  {journal} {\bibinfo  {journal} {Journal of Physics A: Mathematical and
  General}\ }\textbf {\bibinfo {volume} {15}},\ \bibinfo {pages} {3329}
  (\bibinfo {year} {1982})}\BibitemShut {NoStop}%
\bibitem [{\citenamefont {Taroni}\ \emph {et~al.}(2008)\citenamefont {Taroni},
  \citenamefont {Bramwell},\ and\ \citenamefont {Holdsworth}}]{jpcm.20.275233}%
  \BibitemOpen
  \bibfield  {author} {\bibinfo {author} {\bibfnamefont {A.}~\bibnamefont
  {Taroni}}, \bibinfo {author} {\bibfnamefont {S.~T.}\ \bibnamefont
  {Bramwell}}, \ and\ \bibinfo {author} {\bibfnamefont {P.~C.~W.}\ \bibnamefont
  {Holdsworth}},\ }\href {\doibase 10.1088/0953-8984/20/27/275233} {\bibfield
  {journal} {\bibinfo  {journal} {Journal of Physics: Condensed Matter}\
  }\textbf {\bibinfo {volume} {20}},\ \bibinfo {pages} {275233} (\bibinfo
  {year} {2008})}\BibitemShut {NoStop}%
\bibitem [{\citenamefont {Anisimov}\ \emph {et~al.}(1997)\citenamefont
  {Anisimov}, \citenamefont {Aryasetiawan},\ and\ \citenamefont
  {Lichtenstein}}]{jpcm.9.767}%
  \BibitemOpen
  \bibfield  {author} {\bibinfo {author} {\bibfnamefont {V.~I.}\ \bibnamefont
  {Anisimov}}, \bibinfo {author} {\bibfnamefont {F.}~\bibnamefont
  {Aryasetiawan}}, \ and\ \bibinfo {author} {\bibfnamefont {A.~I.}\
  \bibnamefont {Lichtenstein}},\ }\href {\doibase 10.1088/0953-8984/9/4/002}
  {\bibfield  {journal} {\bibinfo  {journal} {Journal of Physics: Condensed
  Matter}\ }\textbf {\bibinfo {volume} {9}},\ \bibinfo {pages} {767} (\bibinfo
  {year} {1997})}\BibitemShut {NoStop}%
\bibitem [{\citenamefont {Pandey}\ and\ \citenamefont
  {Maiti}(2008)}]{prb.78.045120}%
  \BibitemOpen
  \bibfield  {author} {\bibinfo {author} {\bibfnamefont {S.~K.}\ \bibnamefont
  {Pandey}}\ and\ \bibinfo {author} {\bibfnamefont {K.}~\bibnamefont {Maiti}},\
  }\href {\doibase 10.1103/PhysRevB.78.045120} {\bibfield  {journal} {\bibinfo
  {journal} {Phys. Rev. B}\ }\textbf {\bibinfo {volume} {78}},\ \bibinfo
  {pages} {045120} (\bibinfo {year} {2008})}\BibitemShut {NoStop}%
\bibitem [{\citenamefont {Bl\"ugel}\ \emph {et~al.}(1987)\citenamefont
  {Bl\"ugel}, \citenamefont {Akai}, \citenamefont {Zeller},\ and\ \citenamefont
  {Dederichs}}]{prb.35.3271}%
  \BibitemOpen
  \bibfield  {author} {\bibinfo {author} {\bibfnamefont {S.}~\bibnamefont
  {Bl\"ugel}}, \bibinfo {author} {\bibfnamefont {H.}~\bibnamefont {Akai}},
  \bibinfo {author} {\bibfnamefont {R.}~\bibnamefont {Zeller}}, \ and\ \bibinfo
  {author} {\bibfnamefont {P.~H.}\ \bibnamefont {Dederichs}},\ }\href {\doibase
  10.1103/PhysRevB.35.3271} {\bibfield  {journal} {\bibinfo  {journal} {Phys.
  Rev. B}\ }\textbf {\bibinfo {volume} {35}},\ \bibinfo {pages} {3271}
  (\bibinfo {year} {1987})}\BibitemShut {NoStop}%
\bibitem [{\citenamefont {Nov\'ak}\ and\ \citenamefont
  {Chlan}(2010)}]{prb.81.174412}%
  \BibitemOpen
  \bibfield  {author} {\bibinfo {author} {\bibfnamefont {P.}~\bibnamefont
  {Nov\'ak}}\ and\ \bibinfo {author} {\bibfnamefont {V.}~\bibnamefont
  {Chlan}},\ }\href {\doibase 10.1103/PhysRevB.81.174412} {\bibfield  {journal}
  {\bibinfo  {journal} {Phys. Rev. B}\ }\textbf {\bibinfo {volume} {81}},\
  \bibinfo {pages} {174412} (\bibinfo {year} {2010})}\BibitemShut {NoStop}%
\end{thebibliography}%

\end{document}